\documentclass[11pt]{article}
\pdfoutput=1  
\usepackage{graphicx}
\usepackage[english]{babel}
\usepackage{appendix}
\usepackage{latexsym,amsmath,amsfonts,amssymb,booktabs}
\usepackage{verbatim}
\usepackage[usenames,dvipsnames]{color}
\usepackage[font=small]{caption}
\usepackage{slashed,upgreek,amscd,cancel,tensor}
\usepackage{adjustbox}
\usepackage[numbers,compress,square]{natbib}
\usepackage[normalem]{ulem}
\usepackage{upgreek}
\usepackage{stfloats}
\usepackage[stable]{footmisc}
\usepackage{afterpage}
\usepackage{hyperref}
\hypersetup{
pdfencoding=auto
colorlinks=true,
}

\definecolor{MyBlue}{rgb}{0.15,0.15,0.70}

\graphicspath{{./figures/}}
\numberwithin{equation}{section}

\input epsf
\newcommand{\be}{\begin{equation}}
\newcommand{\ee}{\end{equation}}
\newcommand{\beq}{\begin{equation}}
\newcommand{\eeq}{\end{equation}}
\newcommand{\bea}{\begin{eqnarray}}
\newcommand{\eea}{\end{eqnarray}}

\newcommand\alphaB{\alpha_{\text{B}}}

\newcommand{\deltam}{\delta}

\begin{document}
\vspace{0.5cm}

\begin{center}
\Large{\textbf{Constraints on dark energy and modified gravity\\
 from the \textsc{BOSS} Full-Shape and \textsc{DESI} BAO data
}}

\vspace{.5cm}
 P.~Taule$^{1}$, M.~Marinucci$^{2,3}$, G.~Biselli$^{4}$, M.~Pietroni$^{4,5}$,  F.~Vernizzi$^{1}$
\vspace{.5cm}
\\
\small{
\textit{
$^1$Institut de physique th\' eorique, Universit\'e  Paris Saclay \\ [0.05cm] CEA, CNRS, 91191 Gif-sur-Yvette, France  \\
$^2$Dipartimento di Fisica e Astronomia “G. Galilei”, Università degli Studi di Padova, via Marzolo 8, I-35131, Padova, Italy\\
$^3$INFN, Sezione di Padova, via Marzolo 8, I-35131, Padova, Italy\\
$^4$Dipartimento di Scienze Matematiche, Fisiche e Informatiche, Universit\`a di Parma, Parco Area delle Scienze 7/A, I-43124, Parma, Italy \\
$^5$ INFN, Gruppo Collegato di Parma, Parco Area delle Scienze 7/A, I-43124, Parma, Italy
}}
\vspace{.2cm}

\end{center}

E-mail: petter.taule@ipht.fr, marco.marinucci@unipd.it

\vspace{0.5cm}

\begin{abstract}
We constrain dark energy and modified gravity within the effective field theory of dark energy framework using the full-shape BOSS galaxy power spectrum, combined with \textsc{Planck} cosmic microwave background (CMB) data and recent baryon acoustic oscillations (BAO) measurements from DESI. Specifically, we focus on a varying braiding parameter $\alpha_{\rm B}$, a running of the ``effective'' Planck mass $\alpha_{\rm M}$, and a constant dark energy equation of state  $w$. The analysis is performed with two of these parameters at a time, including all the other standard cosmological parameters and marginalizing over bias and nuisance parameters. The full-shape galaxy power spectrum is modeled using the effective field theory of large-scale structure up to 1-loop order in perturbation theory.
We find that the CMB data is most sensitive to $\alpha_{\rm B}$, and that adding large-scale structure information only slightly changes the parameter constraints. However, the large-scale structure data significantly improve the bounds on $\alpha_{\rm M}$ and $w$ by a factor of two.
This improvement is driven by background information contained in the BAO, which breaks the degeneracy with $H_0$ in the CMB.
We confirm this by comparing the \textsc{BOSS} full-shape information with \textsc{BOSS} BAO, finding no significant differences.
This is likely to change with future high-precision full-shape data from \textsc{Euclid} and \textsc{DESI} however, to which the pipeline developed here is immediately applicable.

\end{abstract}

\vspace{1cm}

\newpage
\tableofcontents

\vspace{.5cm}

\section{Introduction}

Testing General Relativity (GR) on cosmic scales and unveiling the nature of Dark Energy (DE) are undoubtedly among the most ambitious goals in cosmology. Specifically, one would like to understand if there are new degrees of freedom  beyond those  posited in the $\Lambda$CDM model, where gravity is mediated by a spin-2 massless field and DE is a cosmological constant.
Building on the legacy of
\textsc{SDSS}~\cite{eBOSS:2020yzd}, ongoing redshift surveys such as \textsc{DESI} \cite{DESI:2016fyo} and
\textsc{Euclid}  \cite{Euclid:2024yrr, Cimatti:2009is} are measuring the large-scale structure of the universe (LSS)  over unprecedentedly large volumes. These data  will enable percent-level determinations of the $\Lambda$CDM
parameters \cite{Euclid:2019clj}, therefore, substantial constraints on a large number of beyond-$\Lambda$CDM scenarios can be expected \cite{Amendola:2016saw}.

From a theoretical perspective, there is no compelling argument favoring a specific  beyond-$\Lambda$CDM scenario  over others. Therefore, model-independent frameworks are essential in order to perform analyses as free from unmotivated theoretical priors as possible. Different model-independent parameterizations of general cosmologies have been proposed (see, for instance~\cite{Creminelli:2008wc, Baker:2011jy, Battye:2012eu,
Gubitosi:2012hu,  Baker:2012zs, Bloomfield:2012ff, Gleyzes:2013ooa,
Bloomfield:2013efa, Battye:2013ida, Gleyzes:2014rba, Skordis:2015yra,
Lagos:2016wyv}). In particular, the Effective Field Theory of Dark Energy (EFTofDE) \cite{Gubitosi:2012hu, Gleyzes:2014rba}
provides a universal description of DE and modified gravity that covers all possible models in which the $\Lambda$CDM degrees of freedom are augmented by a single scalar field. In this symmetry-based approach, it is assumed that the time-diffeomorphism invariance of the gravitational sector is broken due to the presence of the DE field, and the action is expressed in terms of the geometrical quantities
that remain invariant under the preserved spatial diffeomorphisms, multiplied by time-dependent free coefficients. These coefficients encapsulate the complete model-dependence. By appropriately fixing them, any single-field DE model can be matched, as shown explicitly in \cite{Gubitosi:2012hu} in the cases of Quintessence \cite{Zlatev:1998tr}, DGP \cite{Dvali:2000hr}, $k$-essence \cite{Armendariz-Picon:2000nqq}, $f(R)$ gravity \cite{Carroll:2004de} and Horndeski theories \cite{Horndeski:1974wa,Deffayet:2011gz}, among others.

Cosmological observables are natural candidates to look for modified gravity on large-scales.
In the context of the EFTofDE, observational forecasts and bounds have been obtained using the CMB, BAO, $f\sigma_8$, ISW-LSS cross correlation and weak lensing data, see e.g.~\cite{Raveri:2014cka,Gleyzes:2015rua,Bellini:2015xja,Noller:2018wyv, SpurioMancini:2018apc,SpurioMancini:2019rxy,Perenon:2019dpc, Noller:2020afd, Brando:2019xbv,Chudaykin:2024gol} and \cite{Frusciante:2019xia} for a review.
In this paper, we advance these efforts to constrain the landscape of DE models by leveraging the \emph{full-shape galaxy power spectrum} information measured by \textsc{BOSS}~\cite{BOSS:2015npt}, combined with BAO's measured by \textsc{DESI}~\cite{DESI:2024mwx} and CMB (from \textsc{Planck}~\cite{Planck:2018vyg}).
We implement state-of-the art computational schemes for all the relevant observables;
in particular, our computations of the CMB anisotropies and of the full-shape galaxy power spectrum are based on a \emph{consistent treatment of the perturbation equations and of galaxy bias, both at the linear and at the mildly non-linear level}, within the EFTofDE parameterization.
To achieve this, we use the \texttt{hi\_class} code~\cite{Zumalacarregui:2016pph, Bellini:2019syt}, modified to account for the specific models we consider, as detailed below.
Furthermore, we model the full-shape galaxy power spectrum using the EFTofLSS~\cite{Baumann:2010tm, Carrasco:2012cv, Senatore:2014eva, Perko:2016puo} (see reviews in e.g.~\cite{Ivanov:2022mrd, Cabass:2022avo}), which has become a standard framework for LSS analyses in recent years~\cite{Ivanov:2019pdj, DAmico:2019fhj, DAmico:2022gki, Cabass:2022wjy}.
We use \texttt{PyBird}, a fast Python code which computes the 1-loop power spectrum ~\cite{DAmico:2020kxu}, and modify it to capture the altered clustering dynamics in the EFTofDE \cite{Cusin:2017wjg,Cusin:2017mzw,Bose:2018orj}.
The rigorous treatment of modified gravity at the non-linear level to analyze the full-shape \textsc{BOSS} power spectrum represents the main novelty of this work.
Our pipeline is furthermore fully prepared to analyze LSS data from e.g.\ the ongoing \textsc{Euclid}~\cite{Amendola:2016saw} high-precision galaxy survey in the future \cite{Euclid:2023rjj, Euclid:2023tqw,Euclid:2023bgs,Euclid:2024xfd}.

We consider two main scenarios: the first (Model I) involves mixing between scalar and tensor kinetic terms (referred to as \emph{braiding} and generally labeled $\alpha_{\rm B}$) alongside a running ``effective'' Planck mass rate $\alpha_{\rm M}$, with DE equation-of-state $w=-1$.
The second scenario (Model II) features non-zero braiding with a fixed Planck mass and a arbitrary constant equation of state.
We examine two distinct choices for the time-dependence of the EFTofDE parameters, comparing their implications.
In the case of a running Planck mass, we adopt a parameterization which can account for the impact of the running on the background dynamics.
This approach differs from another commonly adopted parameterization in the literature, in which the background is chosen to be unaffected by $\alpha_{\rm M}$, as e.g.\ adopted by \texttt{hi\_class}.
For the LSS analysis, we assume that the dynamics is \emph{quasi-static}, i.e.\ that observed LSS scales are much smaller than the DE sound horizon.
We ensure that this is always the case by tuning the EFTofDE kineticity parameter $\alpha_{\rm K}$ such that the DE sound speed is equal to the speed of light at all times.
Lastly, we discuss the tensor speed excess parameter $\alpha_{\rm T}$.
We find that large-scale structures are only modestly affected by it, and set it to zero in the analysis.

As we will see below, we find that the braiding is mostly constrained by \textsc{Planck} CMB data, and adding LSS full-shape or BAO information only slightly shifts the marginalized posterior probability.
The effective Planck mass running and the DE equation-of-state however, which impact the background at late times, are less strongly constrained by the CMB, due to a degeneracy with $H_0$.
We find that adding LSS information improves the bounds on these parameters: e.g.\ for $\alpha_{\rm M}  \propto a^3$ we obtain for its value today, $\gamma_{\rm M} \equiv \alpha_{\rm M} (a=a_0)$, that $\gamma_{\rm M} < 1.83$ with \textsc{Planck} alone, which is tightened to $\gamma_{\rm M} < 0.964$ by adding the \textsc{BOSS} full-shape power spectrum.
To examine where the enhanced constraining power is coming from, we compare the inference from the full-shape spectrum with that of BAO measurements from BOSS.
We find insignificant differences, indicating that nonlinearities have small impact on the constraints.
This is likely to change when more precise full-shape data from e.g.\ \textsc{Euclid} or \textsc{DESI} will become available.

The paper is structured as follows.
In Sec.~\ref{sec:revEFTofDE}, we review the EFTofDE, define the dimensionless $\alpha$-parameters \cite{Bellini:2014fua} and discuss specific limits in which this approach reproduces well-know dark energy models.
In Sec.~\ref{sec:param} we discuss how we parameterize the background evolution taking into account a running Planck mass and we outline our choice for the functional time-dependence of the $\alpha$-parameters.
Section~\ref{sec:observations} describes the impact on CMB and galaxy clustering observables of this parameterization.
Next, in Sec.~\ref{sec:num_results}, we describe the observational datasets we use and perform the statistical inference on the EFTofDE.
We conclude in Sec.~\ref{sec:Conclusion}.
Appendix~\ref{app:alphaT} discusses the impact on tensor speed excess, illustrating why we expect minimal sensitivity in LSS observables.
Finally, we display full posterior triangle plots and inferences on cosmological parameters in Appendix~\ref{app:full_posteriors}.

\section{Effective field theory of dark energy}
\label{sec:revEFTofDE}

In the EFTofDE framework, it is assumed that the time-dif\-feo\-mor\-phism invariance of the gravitational sector is broken due to the presence of the dark energy field. This assumption allows us to adopt a specific gauge, known as the unitary gauge, where the time aligns with uniform field hypersurfaces. In this gauge, the action can be formulated in terms of geometrical quantities that remain invariant under the preserved time-dependent spatial diffeomorphisms. These include the metric with upper zero indices,  $g^{00}$, the extrinsic curvature of the time
hypersurfaces and its trace, respectively $K_{\mu}^{\nu}$ and $K$, the 3-dimensional Ricci tensor and scalar of these hypersurfaces, respectively $^{(3)}\!R_{\mu}^{\nu}$ and $^{(3)}\!R$, and quantities that are fully diffeomorphism invariant such as the 4-dimensional Ricci scalar $^{(4)}\!R$.

Expanding around a flat Friedman-Lema\^itre-Robertson-Walker (FLRW) background with  metric $ds^2 = -dt^2 + a^2(t)d\vec{x}^2$,  the unitary gauge action takes form~\cite{Gleyzes:2013ooa}
 \begin{equation}
\label{eq:action}
\begin{split}
    S =  \int & \ d^4x\,\sqrt{-g}\bigg[\frac{M_\star^2 f(t)}{2}\, ^{(4)}\!R - \Lambda (t) - c(t) g^{00} + \frac{m_2^4(t)}{2} \left( \delta g^{00} \right)^2  \\
    &- \frac{m_3^3(t)}{2}\delta K \delta g^{00} - m_4^2(t)\left(\delta K^2  - \delta K^\nu_{\mu}\delta K^\mu_{\nu} - \frac{1}{2}\delta g^{00} {}^{(3)}\!R\right)\bigg] \;,
\end{split}
\end{equation}
where $\delta g^{00} \equiv g^{00}+1$, $\delta K  \equiv K  - 3H $ and  $\delta K_{\mu}^{\nu} \equiv K_{\mu}^{\nu} - H \delta_{\mu}^{\nu}$, with $H \equiv \dot a/a$ being the Hubble rate.
To write this action we have restricted to operators appearing in Horndeski theories~\cite{Horndeski:1974wa,Deffayet:2011gz}.\footnote{
The effective approach to cosmological perturbations in beyond Horndeski theories \cite{Gleyzes:2014dya,Zumalacarregui:2013pma}  and in DHOST theories \cite{Langlois:2015cwa} have been developed in \cite{Gleyzes:2013ooa,Gleyzes:2014qga,Gleyzes:2014rba} and \cite{Langlois:2017mxy}, respectively.}
Moreover, we have focused solely on operators that begin at quadratic order, as these directly impact linear perturbations. Indeed, higher-order operators\footnote{Considering only the operators that contribute to the leading number of spatial derivatives in the Horndeski class of theories, there are three more operators that one can write inside the brackets of eq.~\eqref{eq:action}. Two of them  start at cubic order in the perturbations and read~\cite{Cusin:2017wjg,Cusin:2017mzw}
 \begin{equation}
     \begin{split}
         &\delta g^{00} \left(\delta K^2  - \delta K^\nu_{\mu}\delta K^\mu_{\nu} \right) \;, \\
         & \delta K^3 - 3\delta K\delta K^\nu_{\mu}\delta K^\mu_{\nu} + 2\delta K^\nu_{\mu}\delta K^\mu_{\rho}\delta K^\rho_{\nu} + 3 \delta g^{00} \left( \delta K^\nu_\mu - \frac12 \delta g^{00} {}^{(3)}\!R \right) \;.
     \end{split}
 \end{equation}
The third operator starts at quartic order and read
 \begin{equation}
     \delta g^{00} \left( \delta K^3 - 3\delta K\delta K^\nu_{\mu}\delta K^\mu_{\nu} + 2\delta K^\nu_{\mu}\delta K^\mu_{\rho}\delta K^\rho_{\nu} \right)\;.
 \end{equation}
 }
affect only the one-loop  or higher-loops contributions \cite{Cusin:2017wjg,Cusin:2017mzw} and would remain largely unconstrained by the current data.

While $M_\star$ is a constant, the other parameters  exhibit time dependence. Specifically, the parameters $\Lambda$ and $c$ are involved in the background equations and can be expressed in terms of background quantities such as $H$, $f$, their time derivatives, as well as the homogeneous matter energy density $\bar \rho_{\rm m}$ and pressure $\bar p_{\rm m}$. However, the explicit expressions of these parameters will not be necessary for our current discussion.\footnote{
 It is \cite{Gleyzes:2013ooa}
 \begin{equation}
     \begin{split}
         \Lambda &= M_\star^2 \left(\ddot f + 5 \dot f H +2 f \dot H + 6 f H^2 \right)+\frac12 \left( \bar p_{\rm m} - \bar \rho_{\rm m} \right) \;, \\
         c &= M_\star^2 \left(- \ddot f +  \dot f H -2 f \dot H  \right)-\frac12 \left( \bar \rho_{\rm m} + \bar p_{\rm m}  \right) \;.
     \end{split}
     \label{bckgeqs}
 \end{equation}
 }

 To simplify the perturbation equations, it is advantageous to introduce certain combinations of the other parameters. First, we define the time-dependent \textit{effective} Planck mass,
\begin{equation}
    M^2 \equiv M^2_\star f + 2m_4^2\,,
\end{equation}
which sets the normalization of the tensor perturbations of the metric and has to be strictly positive.
Additionally, we introduce dimensionless time-dependent parameters \cite{Bellini:2014fua,Gleyzes:2014rba},
\begin{equation}
\alpha_{\rm K} \equiv \frac{2 c + 4 m_2^4}{M^2 H^2}\;, \quad  \alpha_{\rm B} \equiv \frac{M_\star^2 \dot{f} - m_3^3}{2 M^2 H}\,, \quad \alpha_{\rm M} \equiv \frac{2 \dot M}{M H}\,, \quad \alpha_{\rm T} \equiv -\frac{2m_4^2}{M^2}\;.
\label{eq:alphasdef}
\end{equation}
For the explicit relations connecting these quantities to the functions that appear in the covariant  action of Horndeski theories, we refer the reader to App.~A of \cite{Cusin:2017mzw}.

When expanded at quadratic order in the metric perturbations, eq.~\eqref{eq:action} governs the linear dynamics of scalar and tensor fluctuations (see e.g.~\cite{Gleyzes:2013ooa,Gleyzes:2014rba}). The dispersion relations for these propagating modes are given by $\omega^2 = c_s^2 k^2$ for scalar modes and $\omega^2 = c_{\rm T}^2 k^2$ for tensor modes. The speed of scalar fluctuations is
\begin{equation}
c_s^2 =  \frac{2 \nu}{\alpha} \label{cs}\;,
\end{equation}
where
\begin{align}
 \alpha & \equiv \alpha_{\rm K} + 6 \alpha_{\rm B}^2 \;,
 \label{eq:alpha}\\
 \nu & \equiv  -  (1+\alpha_{\rm B}) \bigg(\frac{\dot H}{H^2} + \xi \bigg) - \frac{\dot \alpha_{\rm B}}{H} -  \frac{\bar \rho_{\rm m} + \bar p_{\rm m}}{2 M^2 H^2} \;, \label{nu}
\end{align}
with
\begin{equation}
    \xi \equiv \alpha_{\rm B} + \alpha_{\rm T} + \alpha_{\rm B} \alpha_{\rm T} - \alpha_{\rm M} \;. \label{eq:xi}
\end{equation}
For tensor modes, $c_{\rm T}^2 = 1+ \alpha_{\rm T}$.

To ensure the absence of gradient instabilities, it is necessary for the speeds of propagation to always be positive. Additionally, the kinetic energy of the scalar mode is proportional to the parameter $\alpha$ defined above, which must be positive to avoid ghost instabilities  \cite{Gleyzes:2013ooa}. In summary, the following four stability conditions are required for these parameters:
\begin{equation}
M \ge 0 \;, \qquad \alpha_{\rm T} \ge -1 \;, \qquad \alpha \ge 0 \;, \qquad c_s^2 \ge 0 \;.
\label{stability}
\end{equation}

Before delving into the observational implications of these theories, let us consider some limiting values of the parameters defined above, in order to better grasp their physical meaning.
First of all, the $\Lambda$CDM limit corresponds to setting
\begin{equation} \nu =0 \;, \qquad \alpha_{\rm B}=\alpha_{\rm M} = \alpha_{\rm T} = 0 \;, \qquad \text{($\Lambda$CDM)}\;. \label{eq:LCDMlimit}
\end{equation} One can check that in this limit the scalar field perturbations decouple from the metric perturbations and one recovers the  perturbation equations of general relativity, independently on the value of $\alpha_{\rm K}$.

With the choice
\begin{equation}
 \nu = - \frac{\dot H}{H^2} - \frac{\bar \rho_{\rm m} + \bar p_{\rm m}}{2 M^2 H^2} \neq 0\,,\qquad \alpha_{\rm B}=\alpha_{\rm M} = \alpha_{\rm T} = 0 \;, \qquad \text{($k$-essence)}\;, \label{eq:kessencelimit}
\end{equation}
with $\alpha = \alpha_{\rm K} \neq 0$ (see Eq.~\eqref{eq:alpha}), the EFTofDE covers  the class of theories known as $k$-essence \cite{Armendariz-Picon:2000nqq},
parameterized by $\alpha$ and $\nu$.
In $k$-essence, DE behaves as an independent component that is gravitationally coupled to the other species, but no genuine modified gravity effects are present.
Note that  since $\nu$ in the above equation does not vanish, in $k$-essence the evolution of the background does not follow the one of the $\Lambda$CDM.
By appropriately choosing $\alpha = 2 \nu$ we recover the  Quintessence scenario \cite{Zlatev:1998tr}, with $c_s^2 = 1$ (see eq.~\eqref{cs}).
Another sub-case of  $k$-essence is  clustering DE, characterized by $c_s^2 \to 0$, which arises for $\alpha  \gg \nu \neq 0$.

The time-dependence of the effective Planck mass is controlled by $\alpha_{\rm M}$, which can also be written as
\beq
\alpha_{\rm M}=\frac{d \log M^2}{d \log a}\,,
\eeq
and, as we will see in the following, is the parameter on which we can extract more information by combining CMB and LSS data.
Furthermore, $\alpha_{\rm B}$ parameterizes the mixing between the scalar and tensor kinetic terms~\cite{Creminelli:2008wc} or {\em braiding} \cite{Deffayet:2010qz}, which represents a genuine modification of gravity.
Notably, it  appears in theories such as Brans-Dicke \cite{Brans:1961sx} and $f(R)$ \cite{Carroll:2004de}, in which case we have $\alpha_{\rm M} = 2 \alpha_{\rm B}$, and in the cubic Galileon \cite{Nicolis:2008in}.
Finally, as explained below Eq.~\eqref{eq:xi}, $\alpha_{\rm T}$ measures deviations of the speed of tensor modes from the speed of light. The observation of the gravitational wave event GW170817 and its electromagnetic counterpart \cite{LIGOScientific:2017vwq} set a stringent bound on such a deviation, of the order of $10^{-15}$ \cite{Creminelli:2017sry,Ezquiaga:2017ekz,Baker:2017hug}, assuming  the regime of validity of the EFTofDE extends to the scales probed by the LIGO/Virgo interferometers  \cite{deRham:2018red}. It is possible to include this parameter in the analysis to obtain an independent constraint using cosmological data. However, as we discuss in App.~\ref{app:alphaT}, large-scale structures are only minimally affected by variations in this parameter. Therefore, in what follows we will fix $\alpha_{\rm T}=0$.

\section{Parameterization}
\label{sec:param}

In this section, we introduce the parameterization adopted for the background quantities and the $\alpha_i$ parameters.

\subsection{Background}
\label{sec:BK}

Let us start discussing how we parameterise the background evolution, in particular the equation of state  $w$, which is
a crucial parameter for observations. For a specific scalar-tensor theory possessing a Lagrangian, the background evolution   is determined by  varying the action with respect to the homogeneous metric and  scalar field. In the EFTofDE framework, where the scalar field has been gauged away,  the  energy density and pressure of dark energy, respectively $\bar \rho_{\rm DE}$ and $\bar p_{\rm DE}$, are not explicitly specified \cite{Gubitosi:2012hu}.%
\footnote{In the EFTofDE action, the parameters $c(t)$ and $\Lambda(t)$ are fully specified by Eqs.~\eqref{bckgeqs} once the time evolution of the Hubble rate, $H = H(t)$, and of $f(t)$, and the energy density and pressure of all other background components have been given. }
This can lead to ambiguities in defining the dark energy background quantities. A natural approach is to use the Friedmann equations, which provides a clear definition in theories where the Planck mass is time-independent, such as Quintessence and $k$-essence. However, in models where
the effective  Planck mass $M$, whose time-dependence is controlled by the parameter $\alpha_{\rm M}$ (see Eq.~\eqref{eq:alphasdef}), varies in time, the standard Friedmann equations no longer apply, because dark energy and curvature are mutually coupled. To avoid ambiguities, in this section we will provide a detailed explanation of how we define the energy density and equation of state for the dark energy component.\footnote{ In this paper, we  adopt a definition of the background energy density and pressure for dark energy that differs from the one used in the public version of the \texttt{hi$\_$class} code. This has required modifications to the public code and complicates direct comparisons with analyses based on the standard \texttt{hi$\_$class} version, such as those by \cite{Noller:2018wyv, Seraille:2024beb, Chudaykin:2024gol}.
With this choice, however, our analyses can account for the impact of a running Planck mass on the background dynamics as well.}

Following \cite{Gleyzes:2014rba},  we will assume that the  {\em modified} Friedmann equations feature a time-dependent effective  Planck mass $M$,
\begin{align}
H^2 &= \frac{1}{3 M^2} (\bar \rho_{\rm m} + \bar \rho_{\rm DE} ) \; ,  \label{Frie1}\\
\dot H &= - \frac{1}{2 M^2}  \left( \bar \rho_{\rm m} +  \bar \rho_{\rm DE} +  \bar p_{\rm DE}  \right) \label{Frie2} \;,
\end{align}
where $\bar \rho_{\rm m}$ is the matter energy density and  the matter pressure is set to zero, $\bar p_{\rm m}=0$.
These equations {\em define}  $\bar \rho_{\rm DE} (t)$ and $\bar p_{\rm DE} (t)$.
From Eq.~\eqref{Frie1}, the energy density fractions of the two components for a critical background are given by
\be
\Omega_{\rm m} (t) \equiv \frac{\bar \rho_{\rm m}}{3 M^2 H^2} \;, \qquad \Omega_{\rm DE} (t) \equiv \frac{\bar \rho_{\rm DE}}{3 M^2 H^2} \;,
\ee
which satisfy $\Omega_{\rm m} (t) + \Omega_{\rm DE} (t)=1$. Finally, using these  equations we can express the equation of state $w \equiv \bar p_{\rm DE}/\bar \rho_{\rm DE}$ in terms of $H^2$, $\dot H$ and  $\Omega_{\rm m}$,
\be
w= - \bigg(1 + \frac23 \frac{\dot H}{H^2} \bigg) (1-  \Omega_{\rm m} )^{-1}\;,
\label{Hdot}
\ee
which we can take as our definition of $w$, valid at late time and assuming flatness.

Let us now turn to the evolution of the background matter energy density.
Our analysis is carried out in the Jordan frame, where matter is minimally coupled to the gravitational metric.
In this frame, the matter stress-energy tensor $T^{(\rm m)}_{\mu \nu}$ is covariantly conserved,
\be
\nabla^\mu T^{(\rm m)}_{\mu  \nu} = 0 \;,
\label{SEcons}
\ee
which implies that $\rho_{\rm m}$ follows the usual conservation equation, i.e.,
\be
\dot {\bar \rho}_{\rm m} + 3 H \bar \rho_{\rm m} =0 \;.
\label{mattercons}
\ee
Consequently, using the above equations and the definition of $\alpha_{\rm M}$, the time evolution of  $\Omega_{\rm m}$  follows from
\be
\dot \Omega_{\rm m} = - H \bigg(  3 + 2 \frac{\dot H}{H^2} +\alpha_{\rm M} \bigg) \Omega_{\rm m}  = H \left[  3 w (1- \Omega_{\rm m}) - \alpha_{\rm M} \right] \Omega_{\rm m}  \;,
\label{eq:Ommdot}
\ee
where the second equality results from Eq.~\eqref{Hdot}.

We can now derive the evolution equation of the dark energy component. By taking the time derivative of Eq.~\eqref{Frie1} and comparing it with Eq.~\eqref{Frie2}, while using Eq.~\eqref{mattercons}, we obtain
\be
\dot {\bar {\rho}}_{\rm DE}  +3 (1+w) H {\bar {\rho}}_{\rm DE} = 3 \alpha_{\rm M}  M^2 H^3 \;,
\label{DEcons}
\ee
which can be integrated  once the functional form of $w = w(t)$ is specified.
In particular, when $\alpha_{\rm M}~\neq~0$ the energy density of dark energy is not constant even for $w = -1$,
meaning that the background evolution deviates from that of $\Lambda$CDM. The appearance of the right-hand side of Eq.~\eqref{DEcons} should not be surprising. In models with a running Planck mass, the dark energy background is non-minimally coupled to gravity, which manifests as a source term in the conservation equation.%
\footnote{To understand this point with a specific example, let us consider a relativistic non-minimally coupled scalar field,
\be
S= \int d^4 x \sqrt{-g} \left[ \frac{M_\star^2 f(\phi) }{2} R - \frac12 (\partial \phi)^2 - V(\phi)  \right]  + S_{\rm m}\;,
\ee
where $S_{\rm m}$ is the matter action.
(The EFTofDE action reproducing this model is obtained by choosing $c(t)= \dot {\bar \phi}^2 (t)/2$, $\Lambda (t) = V(\bar \phi(t))$ and $m_i(t)=0$ in Eq.\eqref{eq:action}.)
By varying with respect to the metric, one obtains
\be
f(\phi) G_{\mu \nu} =  \frac1{M_\star^2} T_{\mu \nu}^{(\rm m)} + \frac1{ M_\star^2} T_{\mu \nu}^{(\phi)} \;,
\ee
where the  matter stress-energy tensor $T_{\mu \nu}^{(\rm m)} $ is  defined in the standard way, while $T_{\mu \nu}^{(\phi)}$ is an {\em effective} dark energy stress-energy tensor, defined as
\be
 T_{\mu \nu}^{(\phi)} = \partial_\mu \phi \partial_\nu \phi - \frac12 g_{\mu \nu} \left( \frac12 (\partial \phi)^2 + V(\phi)\right) + \nabla_\mu \nabla_\nu f(\phi) - g_{\mu \nu} \square f(\phi)  \;.
\ee
(Note that this definition is a matter of convention and other choices are possible.)
Using the Bianchi identity and the conservation of $T_{\mu \nu}^{(\rm m)} $, Eq.~\eqref{SEcons}, one obtains
\be
\nabla^\mu  T_{\mu \nu}^{(\phi)} = M_\star^2 f'(\phi) \nabla^\mu \phi G_{\mu \nu} \;,
\ee
which in the homogeneous limit reduces to Eq.~\eqref{DEcons}, with $\bar \rho_{\rm DE} =  \bar {T}^{(\phi)}_{00}$ and $\bar p_{\rm DE} =  a^{-2} \bar {T}^{(\phi)}_{ii}$.}

\subsection{Perturbations}
\label{sec:code}

The evolution of perturbations is fully determined by the time-dependent parameters  $ f(t)$ and $m_i(t)$ in the action~\eqref{eq:action} or, equivalently, by the dimensionless parameters $\alpha_i(t)$. The time evolution of these parameters depends on the specific model. In  covariant models, these parameters can be linked to the functions in the Lagrangian, with their time evolution dictated by the scalar field evolution. Here, we adopt a phenomenological approach, fixing the time evolution of the $\alpha_i$ in a convenient way, with the expectation that the approach presented here can be easily extended to other parameterizations.

Given our focus on analyzing genuine modifications of gravity, we consistently include the braiding parameter $\alpha_{\rm B}$
in our posterior distributions. Additionally, since the available data cannot robustly constrain more than two dark energy parameters simultaneously, we consider combining $\alpha_{\rm B}$ in pairs with either $w$ or $\alpha_{\rm M}$.  We only consider a constant $w$ and parameterize the time-dependence of $\alpha_{\rm B}$ and $\alpha_{\rm M}$ as a function of the scale factor $a$, as%
\footnote{In general, $\alpha_{\rm B}$ and $\alpha_{\rm M}$ are expected to have different time dependencies. While our analysis could in principle account for different exponents for $\alpha_{\rm B}$ and $\alpha_{\rm M}$, here  we have chosen to use a common exponent $p$ for both parameters to simplify the analysis given the available data.}
\begin{equation}
\label{Gammapar}
    \alpha_{\rm B} (a) = \gamma_{\rm B} \,\left( \frac{a}{a_0} \right)^p \,, \qquad
    \alpha_{\rm M}  (a) = \gamma_{\rm M} \, \left( \frac{a}{a_0} \right)^p \;,
\end{equation}
where $\gamma_{i} = \alpha_i(a=a_0)$ are constants and we focus on two specific values of $p$: $p=3/2$ and $p=3$.
Notably, the choice $p=3 $ approximates another common parameterization studied in the literature, where $\alpha_{i} \propto \Omega_{\rm DE}$~\cite{Bellini:2014fua}.
One could generalize this ansatz by letting $p$ be a free parameter in the analysis. However, previous studies have shown that $p$ is largely unconstrained with Planck CMB data~\cite{Noller:2018wyv}.
We do not expect the two redshift bins of the BOSS data to significantly improve the constraining power and, therefore, we consider $p$ fixed in the analysis.
See also \cite{Linder:2015rcz,Linder:2016wqw,Gleyzes:2017kpi,Denissenya:2018mqs,Lombriser:2018olq,Traykova:2021hbr}  for further discussions on EFTofDE parameterizations.

As discussed in Sec.~\ref{sec:revEFTofDE}, we will set $\alpha_{\rm T} =0$. Applying furthermore eqs.~\eqref{Hdot} and \eqref{Gammapar} to the definition of $\nu$, Eq.~\eqref{nu}, we can provide a simplified expression for this parameter, useful for our discussion:
\be
 \nu =  (1+ \alpha_{\rm B})\left[   \frac32  ( 1 + w )( 1-  \Omega_{\rm m} )  - \alpha_{\rm B}+\alpha_{\rm M}  \right]  + \left(  \frac32 \Omega_{\rm m} -p  \right) \alpha_{\rm B}  \;. \label{nunew}
\ee
Stability (i.e.\ no ghost instabilities, $\alpha > 0$ in eq.~\eqref{stability})
requires that this quantity is positive.

The remaining parameter is $\alpha_{\rm K}$. As explained above, the framework developed in Sec.~\ref{sec:DM} requires to work on spatial scales that are smaller than the sound horizon of dark energy, or equivalently for wavenumbers $k \gg a H/c_s$. These are the scales where the quasi-static approximation holds. As shown in \cite{Sawicki:2015zya}, this approximation should be reliable for current surveys as long as the sound speed exceeds 10\% of the speed of light, i.e.~$c_s \gtrsim 0.1$. To ensure that this approximation is always satisfied during the whole cosmic history and, in particular, to prevent entry into a regime of clustering dark energy at late times, we will consider models with a speed of scalar fluctuations equal to the speed of light, i.e., $c_s^2=1$.  By using equation \eqref{cs}, this corresponds to choosing the parameter $\alpha_{\rm K}$ to satisfy the relation
\begin{equation}
\alpha_{\rm K} = 2 \nu - 6 \alpha_{\rm B}^2 \,,
\end{equation}
at any time.
We expect that constraints derived in this manner from large-scale structure data are also applicable to other values of $c_s^2$ (and therefore  of $\alpha_{\rm K}$), as long as $c_s \gtrsim 0.1$.

Tachyon instabilities ensue when the mass squared of the perturbations become negative.
As tachyon instabilities may lead to DE isocurvature modes which dominate over the adiabatic one in the early Universe, we opt to exclude those instabilities, adopting the default settings of \texttt{hi\_class} \cite{Bellini:2019syt}.

In summary, in the following we will focus on two specific models: Model I, where the parameters varied in the analysis are  $\gamma_{\rm B} $ and $ \gamma_{\rm M}$, while $w=-1$ and $\alpha_{\rm T}=0$; Model II, where the parameters varied in the analysis are  $\gamma_{\rm B} $ and $ w$, while $\alpha_{\rm M} = \alpha_{\rm T}=0$.

\section{Link with the observables}
\label{sec:observations}

Here we discuss how the parameterization discussed above enters the relevant perturbation equations and examine its impact on the observables.

\subsection{Dark matter perturbations}
\label{sec:DM}

We use a modified version of \texttt{PyBird}~\cite{DAmico:2020kxu} to compute the impact of dark energy and modified gravity on the galaxy power spectrum in redshift space. Here, we review how the $\alpha_i$ parameters are incorporated into the perturbation equations that describe the long-wavelength evolution of dark matter. For a detailed description of how these parameters propagate to the redshift space power spectrum and their implementation in \texttt{PyBird}, we refer the reader to \cite{Piga:2022mge}.

To study the  impact of dark energy and modified gravity described above on perturbations, we concentrate on scales much shorter than the Hubble radius, where relativistic effects due to the expansion of the universe can be neglected. Moreover, we consider non-relativistic gravitational fields and velocities. For gravitational and field fluctuations below the scalar field sound horizon, we can assume the quasi-static approximation and time derivatives can be taken to be much smaller than spatial derivatives.
For concreteness,
we focus on scalar perturbations  in Newtonian gauge, for which the metric can be written as
\be
ds^2 = - (1+ 2 \Phi) dt^2 + a^2(t) (1- 2 \Psi) d{\bf x}^2 \;.
\ee

Let us start by discussing the impact of dark matter clustering. In the Jordan frame, dark matter particles follow geodesics and the behavior of the dark matter fluid is described by the standard continuity and Euler equations. We define $\delta \equiv \rho_{\rm m}/\bar \rho_{\rm m}-1$ as the dark matter density contrast, where $\rho_{\rm m}$ represents the inhomogeneous dark matter energy density. Additionally, $v^i$ denotes the velocity of dark matter. The continuity  and Euler equation read
\begin{align}  \label{finalconteq}
 \dot \delta + a^{-1} \partial_i \left( (1+ \delta ) v^i \right) & = 0 \; , \\
  \dot v^i + H v^i + \frac1{a} v^j \partial_j v^i   + \frac{1}{a}  \partial_i \Phi  & = - \frac{1}{a \rho_{\rm m}} \partial_j \tau^{ij} \; ,    \label{finaleulereq} \end{align}
where a dot denotes a time derivative. We have employed the smoothed continuity and Euler equations, following the EFT approach proposed in \cite{Baumann:2010tm, Pietroni:2011iz, Carrasco:2012cv}. The right-hand side of Eq.~\eqref{finaleulereq} represents the effective stress-energy tensor, which accounts for the influence of short modes on the dynamics of long modes resulting from the  smoothing procedure. For a  treatment of this smoothing procedure and the stress-energy tensor in the context of modified gravity models, we refer  to \cite{Cusin:2017wjg}.

In order to close these equations, it is necessary to establish a relationship between the gravitational potential $\Phi$ and the dark matter density contrast $\delta$. In general relativity, this is accomplished through the Poisson equation. However, in Horndeski theories, the dynamics of the scalar field must also be taken into account alongside gravitational dynamics. We make the assumption that the mass of the scalar field responsible for modifying gravity is much smaller than the fundamental frequency of the survey and can be neglected. Furthermore, we adopt the quasi-static approximation, neglecting time derivatives. By employing the complete field equations involving both $\Phi$ and $\Psi$, as well as the scalar field equation, it becomes possible to solve for the Laplacian of $\Phi$ in terms of the density contrast and its spatial derivatives. This procedure is  explained in \cite{Cusin:2017mzw, Cusin:2017wjg}. The resulting expression encompasses a term linear in $\delta$, similar to the Poisson equation, but also includes higher-order terms in general. Since our aim is to compute the 1-loop power spectrum, we will only consider terms up to third order in $\delta$.
In this case, we obtain
\begin{align}
\label{sol_NL1}
\frac{\partial^2 \Phi}{H^2 a^2} = & \   \frac{3 \, \Omega_{{\rm m}} }{2}  \,  \mu_{\Phi}(t) \,  \deltam+  \left( \frac{3\, \Omega_{{\rm m}}}{2}  \right)^2 \mu_{\Phi,2}(t)  \left[\deltam^2-\left( \frac{\partial_i\partial_j \deltam}{\partial^{2}}\right)^2\right] \nonumber \\
&+\left(  \frac{3\, \Omega_{{\rm m}}}{2} \right)^3 \mu_{\Phi,22} (t) \left[\deltam-    \frac{\partial_i\partial_j \deltam}{\partial^2}    \frac{\partial_i\partial_j}{\partial^2} \right]\left[\deltam^2-\left(\frac{\partial_k\partial_l \deltam}{\partial^{2}} \right)^2\right]   \nonumber  \\
&+\left(  \frac{3\, \Omega_{{\rm m}}}{2}  \right)^3 \mu_{\Phi, 3} (t) \left[\deltam^3-3\deltam\left( \frac{\partial_i\partial_j  \deltam}{\partial^{2}}\right)^2+2 \frac{\partial_i\partial_j \deltam}{\partial^{2}}  \frac{\partial_k\partial_j \deltam }{\partial^{2}}  \frac{\partial_i\partial_k \deltam}{\partial^{2}} \right]    + {\cal O} (\delta^4)\,.
\end{align}
The other time-dependent quantities are defined in terms of the $\alpha_i$ parameters  as
\begin{align}
    \mu_\Phi & = 1  + \frac{(\alpha_{\rm B}-\alpha_{\rm M})^2}{\nu}\;, \\
    \mu_{\Phi,2} & =-2  \frac{(\alpha_{\rm B}-\alpha_{\rm M})^3}{4 \nu^3}(2\alpha_{\rm B} - \alpha_{\rm M})\,,\\
        \mu_{\Phi,22} & =    \frac{(\alpha_{\rm B}-\alpha_{\rm M})^4}{2\nu^5}  (2 \alpha_{\rm B} -\alpha_{\rm M} )^2 \,,\\
        \mu_{\Phi,3} &= 0 \,,
        \label{eq:nonlinDE}
\end{align}
where we remind the reader that
we have set $\alpha_{\rm T} = 0$. It can be readily verified that when $\alpha_{\rm B} = \alpha_{\rm M}  = 0$, corresponding to the $k$-essence scenario, $\mu_{\Phi}=1$ and  $\mu_{\Phi,2}=\mu_3=\mu_{22}=0$ and the standard Poisson equation is recovered. Note also that in Brans-Dicke and $f(R)$ gravity, for which $\alpha_{\rm M} - 2 \alpha_{\rm B} = m_3^3/(M_\star^2 H) =0 $, we have $\mu_{\Phi,2} = 0 = \mu_{\Phi,22}$.\footnote{In deriving Eq.~\eqref{sol_NL1}, we have assumed that the relevant scales are much shorter than the Compton wavelength of the scalar field. In particular, at larger scales, $\mu_{\Phi,2}$ and $\mu_{\Phi,22}$ do not vanish when $\alpha_{\rm M} = 2\alpha_{\rm B}$. For a detailed discussion on scale dependence in the kernels at larger scales, see, e.g., \cite{Taruya:2014faa}.}

\subsection{Cosmic microwave background }
\label{sec:CMB}

To study the impact of dark energy and modified gravity on CMB anisotropies, a relativistic Einstein-Boltzmann solver that includes these effects within the EFTofDE framework is required. In this work, we use a modified version of \texttt{hi\_class} \cite{Bellini:2019syt}. Other Einstein-Boltzmann codes incorporating modifications of gravity in the EFTofDE approach exist; see e.g.~\cite{Hu:2013twa, Raveri:2014cka, Huang:2012mt} and \cite{Bellini:2017avd} for a code comparison.

While the $\alpha_i$ parameters enter the relativistic perturbation equations in a complex way, we can estimate some of their main effects on the CMB anisotropies, in particular on the Integrated Sachs-Wolfe (ISW) effect and the CMB lensing, using simplified formulas derived in the quasi-static approximation. Specifically, the impact of dark energy and modified gravity on photon geodesics is mediated through the scalar Weyl potential $\Phi+\Psi$. In the quasi-static regime, the relation between this potential and the dark matter density perturbations is given by \cite{Gleyzes:2015rua,DAmico:2016ntq}
\be
\label{Weylpot}
\partial^2 ( \Phi  + \Psi) =
  \   \frac{3 \, \Omega_{{\rm m},t} }{2}  H^2 a^2  \mu_{\rm WL}(t) \,  \deltam \;,
\ee
with
\be
\mu_{\rm WL} = 2  + \frac{(\alpha_{\rm B} -\alpha_{\rm M})(2 \alphaB - \alpha_{\rm M}) }{ \nu}  \;.
\ee

The CMB lensing potential is determined by integrating the Weyl potential along the line of sight, weighted by a window function:
\be
\phi (\hat n)= \int^{z_*}_0  \frac{d z}{H(z)} \frac{\chi(z_*) - \chi(z) }{\chi(z)\chi(z_*)} \left[ \Phi(\chi \hat n, z) + \Psi(\chi \hat n, z)  \right] \;,
\ee
where $\chi \equiv \int_0^z dz /H(z)$ is the conformal distance and $z_*$ denotes the redshift of the last scattering.

On the other hand, the ISW effect depends on the variation of the Weyl potential along the line of sight, i.e.,
\be
\frac{\Delta T}{T} (\hat n) = - \int^{z_*}_0 dz \, \left[ \partial_z \Phi(\chi \hat n, z) + \partial_z  \Psi(\chi \hat n, z)  \right] \;.
\ee
To estimate the effects of modified gravity on the ISW, we can take the derivative with respect to the redshift of Eq.~\eqref{Weylpot}. This yields the following expression, which only holds in the quasi-static limit,
\be
\partial_z \Phi(\chi \hat n, z) + \partial_z  \Psi(\chi \hat n, z) = \frac{1}{1+z}\left( 1- f_{\rm QS} - \frac{d \ln \mu_{\rm WL}}{d \ln a}  \right)  \left[ \Phi(\chi \hat n, z) +   \Psi(\chi \hat n, z) \right] \;,
\label{eq:ISW_impact}
\ee
where
\be
f_{\rm QS} \equiv \left. \frac{d \ln \delta}{d \ln a} \right|_{\rm QS} \;
\ee
is the growth rate computed using the quasi-static approximation.

\subsection{Observable impact}

Finally, we end this section by a summery of the general impact of modified gravity on cosmological observables within
the EFTofDE framework, discussing first the CMB anisotropies, then the BAO and finally the full-shape galaxy power spectrum.

\begin{figure}[t]
    \centering
    \includegraphics[width=0.75\linewidth]{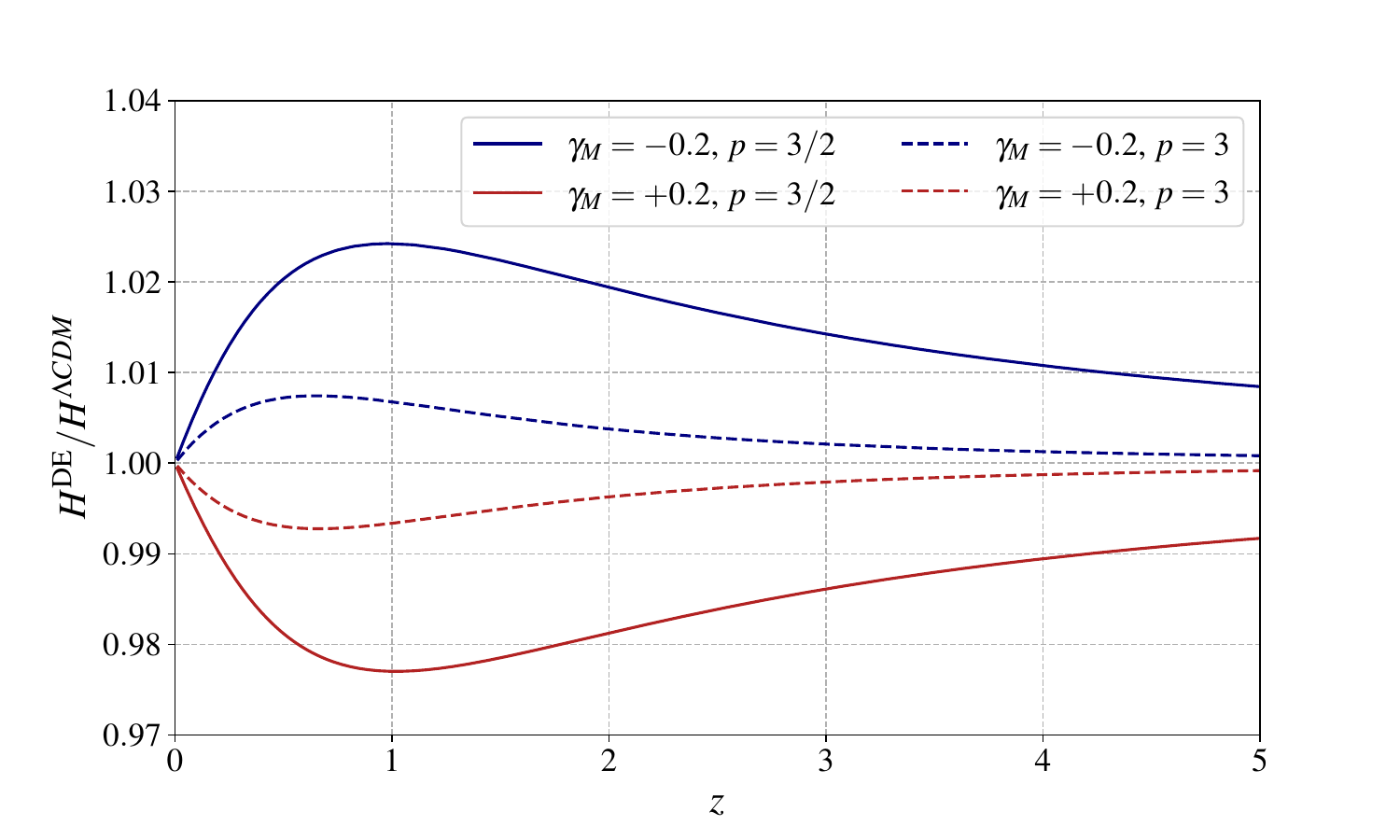}
    \caption{Hubble rate as a function of redshift in models with running Planck mass with $p=3/2$ (solid) and $p=3$ (dashed).  }
    \label{fig:Hplot}
\end{figure}
At the perturbative level, the CMB is sensitive to  $\alpha_{\rm B}$ as well as  $\alpha_{\rm M}$ mainly through the late-ISW effect.
The impact on the ISW  is given by Eq.~\eqref{eq:ISW_impact}, from which it follows that a positive $\alpha_{\rm M}$ slightly enhances the low-$\ell$ spectrum while a non-zero $\alpha_{\rm B}$ lowers the low-$\ell$'s until around $\alpha_{\rm B} \lesssim -0.1$, where its impact on the ISW becomes large and negative, leading to an overall increase.
In addition, $\alpha_{\mathrm{M}}$ impacts the background by sourcing the dark energy density, see Sec.~\ref{sec:BK}.
A demonstration of this effect is displayed in Fig.~\ref{fig:Hplot}, where we show the late-time Hubble rate as a function of redshift in two models with running Planck mass, relative to the $\Lambda$CDM one.
For $p=3$ (dashed line) we see less than a percent difference for $\gamma_{\rm M} = \pm 0.2$, while for $p=3/2$ (solid line), in which the modifications to gravity are important at even earlier times, the difference exceeds 2\%.
Therefore, in addition to the late-ISW impact, $\alpha_{\rm M}$ also affects the location of the acoustic peaks through the angular diameter distance.
This effect can be compensated for by a change of the Hubble rate today, and thus we expect a strong $\gamma_{\rm M}$--$H_0$ degeneracy.

\begin{figure}[t]
    \centering
    \includegraphics[width=0.48\linewidth]{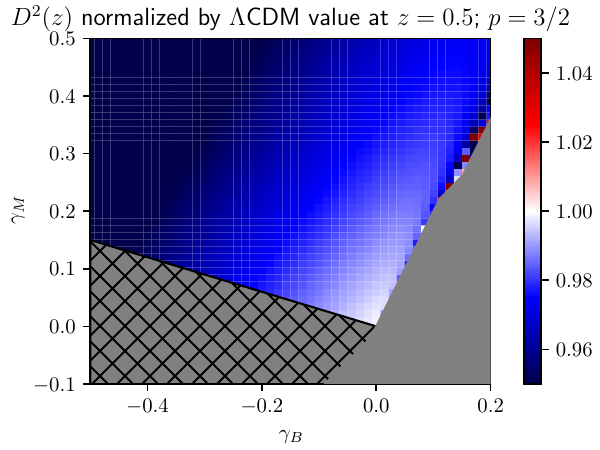}
    \includegraphics[width=0.48\linewidth]{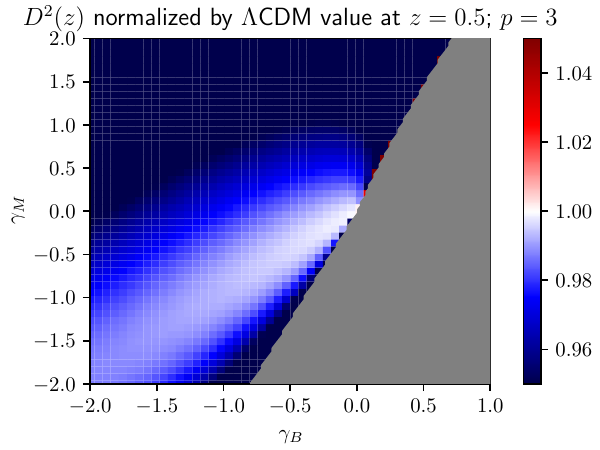}
    \includegraphics[width=0.48\linewidth]{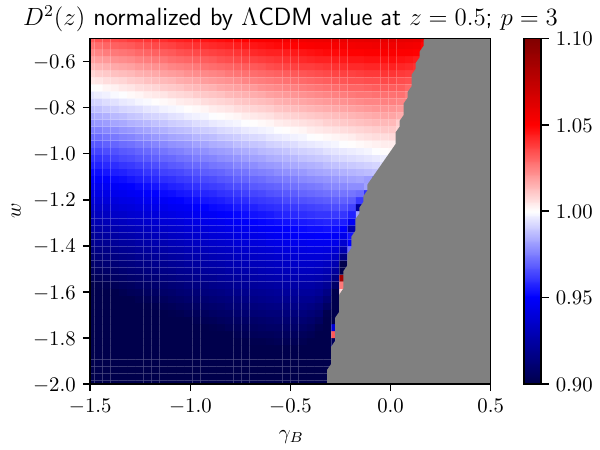}
    \caption{Growth factor $D^2$ at $z=0.5$ compared to the $\Lambda$CDM value in the different models.
    The comparison is performed normalizing $D(z=0) = 1$.
    The gray shaded regions indicate ghost instabilities, while the lighter gray hashed region show tachyonic instabilities.
    \emph{Upper left}: Model I with $p=3/2$. \emph{Upper right}: Model I with $p=3$. \emph{Lower}: Model II with $p=3$.}
    \label{fig:D_contour}
\end{figure}
In this work we include BAO measurements from both the \textsc{SDSS} and \textsc{DESI} surveys.
BAO alone cannot constrain the braiding $\alpha_{\rm B}$, which only affects perturbations,
but it can test background dynamics sourced by the running Planck mass (Model
I) or a dark energy equation  of state $w \neq -1$ (Model II).

Finally, the focus on this work is to test modified gravity with full-shape information
from the \textsc{BOSS} redshift survey. In  modified gravity, galaxy formation
follows the modified Poisson equation~\eqref{sol_NL1}. At the linear level, the growth
factor $D(z)$ is altered by $\mu_{\Phi}(t)$. In Fig.~\ref{fig:D_contour}, we display
it as a function of the EFTofDE parameters in the different models at $z=0.5$.
For non-zero braiding and running of the effective Planck mass (Model I), there is a clear degeneracy
in the $\gamma_{\rm B} \simeq \gamma_{\rm M}$ direction, which was anticipated in the discussion below Eq.~\eqref{eq:nonlinDE}. Similarly, we see a degeneracy
direction in the $\gamma_{\rm B}$--$w$ plane. Beyond the linear level, the $\mu_{\Phi,2}$- and
$\mu_{\Phi,22}$-functions leave imprints on the power spectrum 1-loop correction.%
\footnote{$\mu_{\Phi,3}(t)$ does not enter the 1-loop power
spectrum~\cite{Cusin:2017wjg} and, moreover, it is zero in all the models considered in this work; see Eq.~\eqref{eq:nonlinDE}.}
Lastly, galaxy clustering data is also sensitive to modification of the background via the Alcock-Paczynski effect.

For the galaxy power spectrum in real space, a modification of the growth
factor at a specific redshift can be compensated for by a change of the initial
amplitude of fluctuations as well as by the linear bias. On the other hand,
redshift space distortions are impacted by the modified growth and furthermore
observations at multiple redshift can aid in breaking degeneracies. All in all,
we expect that \textsc{BOSS} power spectrum alone cannot sufficiently test a
two-parameter extension within the EFTofDE, due to strong correlations with
other cosmological parameters, in particular $A_s$ and $n_s$. This is in line
with previous studies on modified gravity with the BOSS
full-shape~\cite{Piga:2022mge, Moretti:2023drg}. Therefore it is useful to
combine the \textsc{BOSS} full-shape with probes that well constrain $A_s$ and $n_s$,
such as CMB measurements, as we will do below.

\section{Results}
\label{sec:num_results}
In this section we list the datasets we use and perform the statistical inference on the EFTofDE models using cosmological data.

\subsection{Datasets}
\label{sec:datasets}
In our analysis we  combine a number of early and late time probes:
\begin{itemize}
\item[\textbf{CMB}:] Our CMB dataset comprises \textsc{Planck} \texttt{TT,TE,EE} high-$\ell$ data as well as low-$\ell$ \texttt{TT} and \texttt{EE} data~\cite{Planck:2018vyg}. We use the official PR3 likelihood for the statistical analysis~\cite{Planck:2019nip}. This dataset is referred to as P18.
\item[\textbf{LSS}:] We include the analysis of the full-shape power spectrum of luminous red galaxies (LRG) from the twelfth data release of the \textsc{BOSS} survey~\cite{BOSS:2015npt}. We analyze the monopole and quadrupole of the redshift-space power spectrum for both North and South Galactic Caps for each redshift bin, $z = 0.32, 0.57$. Following~\cite{Piga:2022mge}, we include modes up to $k_{\rm max} = 0.20\, h$/Mpc for the lower redshift bin and $k_{\rm max} = 0.23\, h$/Mpc for the higher one. This dataset is referred to as FS.

\item[\textbf{BAO}:] To compare the constraining power of the full-shape power spectrum with BAO measurements in the \textsc{BOSS} dataset, we use the \textsc{BOSS} DR7 MGS (Main Galaxy Sample)~\cite{Ross:2014qpa} of galaxies in $z \in [0.07, 0.2]$ as well as the \textsc{BOSS} DR12~\cite{BOSS:2016wmc} samples of galaxies in $z \in [0.2,0.5]$ and $z \in [0.4,0.6]$. We will refer to this dataset as \textsc{BOSS} BAO.
We combine these with the recently released \textsc{DESI} DR1 BAO data, after excluding the two redshift bins that overlap with the \textsc{BOSS} ones. They consist of the LRG2 sample in $0.6 < z < 0.8$, the combined LRG3+ELG1 sample in $0.8 < z < 1.1$, the ELG2 sample in $1.1 < z < 1.6$, the quasar sample in $0.8 < z < 2.1$ and the Lyman-$\alpha$ Forest sample in $1.77 < z < 4.16$. We will refer to this dataset as DESI$^*$.
\end{itemize}

When analyzing the data we always include the effects of neutrinos in the linear evolution, adopting the \textsc{Planck} prescription of one massive neutrino species with $m_\nu = 0.06$ eV and two massless ones, with $N_{\rm eff} = 2.0328$~\cite{Planck:2018vyg}. The theory models are evaluated using a modified version of \texttt{hi\_class}
which accounts for our parameterization and modified background dynamics as described in Secs.~\ref{sec:BK} and \ref{sec:code}.

For the full-shape power spectrum computation we use \texttt{PyBird} \cite{DAmico:2019fhj,DAmico:2020kxu}, accordingly modified to include the linear and mildly non-linear effects of the EFTofDE model \cite{Cusin:2017mzw,Cusin:2017wjg}. The modifications are analogous to those of~\cite{Piga:2022mge} and we
refer the reader to this reference for details on the implementation.
Specifically, following \cite{DAmico:2019fhj}, we use the following set of nuisance parameters:
\be
\{b_1, c_2, b_3, c_{ct}, c_{r,1}, c_{\epsilon,0}, c_{\epsilon,1}, c_{\epsilon,2}\}\,,
\ee
where $b_1$ is the linear bias, $c_2$ is a combination of bias parameters appearing at second order, and $b_3$ is a bias parameter entering the third-order bias expansion.
Additionally, $c_{ct}$ and $c_{r,1}$ are counterterms in redshift space, while $c_{\epsilon,0}$, $c_{\epsilon,1}$ and $c_{\epsilon,2}$ are stochastic terms.
Since we are not sensitive to it, we set to zero the remaining combination of bias parameters appearing at second-order.
We analytically marginalize on the third order bias $b_3$ and on all the counterterms and shot noise ones, following~\cite{DAmico:2020kxu}, using the same priors presented in~\cite{Piga:2022mge}.

The choice of priors on bias parameters in full shape analyses of the BOSS data has been shown to yield significant projection effects in particular on $A_s$ and $n_s$ in $\Lambda$CDM~\cite{Simon:2022lde}.
However, increasing the observational volume or including CMB observations removes these~\cite{Simon:2022lde}.
Similarly, Ref.~\cite{Carrilho:2022mon} analyzed the impact on prior choices for a dark energy model with two extra parameters, finding that projection effects go away when CMB information is included.
In the data analysis below, we check that the posterior mean is not notably shifted compared to the best-fit values (with analytical marginalization).
Given that all our analyses include Planck CMB data, we do not expect significant prior volume effects from the marginalized bias parameters.

We run MCMCs using the public code \texttt{MontePython 3}~\cite{Audren:2012wb, Brinckmann:2018cvx}, varying the cosmological parameters
\begin{equation}
    \{\omega_b, \omega_{cdm}, H_0, A_s, n_s, \tau_{\mathrm{reio}}\}\,,
\end{equation}
in combination with $\{\gamma_{\rm B}, \gamma_{\rm M}\}$ for Model I and $\{\gamma_{\rm B}, w\}$ for Model II, all with flat priors.
We state convergence of the chains when the Gelman-Rubin $R-1$ value~\cite{Gelman:1992zz} is lower than $0.02$.

\subsection{Parameter constraints}
\label{sec:constraints}

In this section, we present the results of the statistical inference on the
EFTofDE. We will address the two models defined in Sec.~\ref{sec:code}.
As for the time dependence, for Model II we analyze only the case $p=3$,
as $p=3/2$ leads to tachyonic instabilities across the entire parameter space, except the $\Lambda$CDM limit.

\begin{figure}[t]
    \centering
    \includegraphics[width=0.49\linewidth]{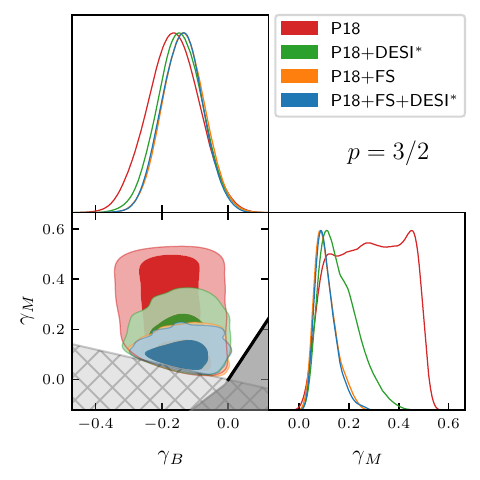}
    \includegraphics[width=0.49\linewidth]{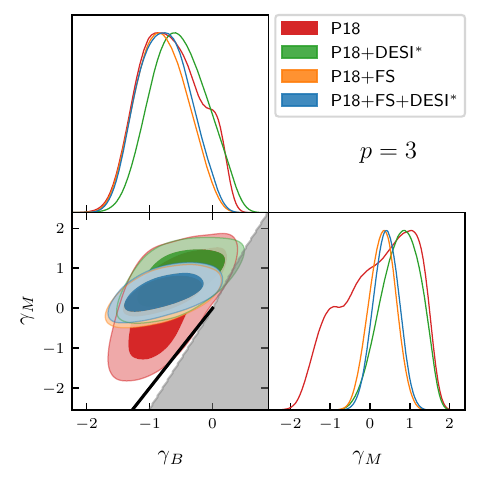}
        \caption{Posterior probabilities for Model I ($\gamma_{\rm B}$ and $\gamma_{\rm M}$
        non-zero) with $p=3/2$ (left) and $p=3$ (right). The different dataset
        combination are defined in Sec.~\ref{sec:datasets}; note that DESI$^*$ refers to a subset of the \textsc{DESI} BAO measurements in which the lowest redshift bins are removed, see that section for details.
        Regions of parameter space which lead to ghost or tachyonic instabilities are indicated by darker gray and lighter gray hashed shading, respectively.
        The region of parameter space which corresponds to Brans-Dicke gravity is indicated by the black line.
        }
    \label{fig:gamma_BM_2x2}
\end{figure}
Figure~\ref{fig:gamma_BM_2x2} shows marginalized posterior distributions on the
EFTofDE parameters in Model I, with $p=3/2$ (left panel) and $p=3$ (right panel).
We find no significant evidence for modified gravity in any of the parameterizations.%
\footnote{For $p=3/2$, $\gamma_{\mathrm{B}} = \gamma_{\mathrm{M}} = 0$ is outside the 2$\sigma$ posterior contour.
This is due to the tachyonic instability region limiting the available parameter space around the $\Lambda$CDM point in combination with the slight preference in the CMB data for non-zero $\gamma_{\mathrm{B}}$.
Hence we do not interpret this effect as significant evidence for modified gravity.}
Additionally, there are no strong correlations between the EFTofDE parameters and the standard cosmological parameters, except for between $\alpha_{\rm M}$ and $H_0$ (the full triangle plot, including the standard cosmological parameters, is included in Appendix~\ref{app:full_posteriors}).
Using \textsc{Planck} alone, these parameters are almost perfectly degenerate because $\gamma_{\rm M} > 0$ leads to a larger $H(z)$ at late times and vice versa.
Adding late-time information on $H(z)$ from full-shape \textsc{BOSS} and/or \textsc{DESI} BAO data breaks this degeneracy, thus tightening constrains on $\gamma_{\rm M}$.
Apart from $H_0$, find no significant shifts of the cosmological parameters in the presence of modified gravity compared to $\Lambda$CDM (see Appendix~\ref{app:full_posteriors}).
For $p=3/2$, $\gamma_{\rm M} < 0$ leads to tachyonic instabilities when $\gamma_{\rm B} < 0$ and ghost instabilities when $\gamma_{\rm B} > 0$, excluding these regions from our analysis.
We find that $\gamma_{\rm M} < 0.508$ at 2$\sigma$ for $p=3/2$ with \textsc{Planck} alone, improved to $\gamma_{\rm M} < 0.203$ after adding the full-shape power spectrum, showing that adding LSS data shrinks the errorbars on this parameter by a factor of 2.
Similarly, for $p=3$ we can constrain $\gamma_{\rm M} < 1.83$ at 2$\sigma$ with \textsc{Planck} alone, updated to $\gamma_{\rm M} < 0.964$ after adding the full-shape.
We report the measured mean and 1$\sigma$ uncertainty from the statistical inference on the EFTofDE parameters and the Hubble constant in Table~\ref{tab:measurements}.

Next, we examine constraints on the braiding coefficient $\gamma_{\rm B}$.
In all models there is a slight preference for a non-zero $\gamma_{\rm B}$ (e.g.\ for Planck+FS+DESI$^{*}$ we obtain $\gamma_{\rm B} < -0.015$ and $\gamma_{\rm B} < -0.026$ at 2$\sigma$ for $p=3/2$ and $p=3$, respectively), primarily driven by low-$\ell$ CMB data.
This preference has also been reported by other studies using CMB data, e.g.~\cite{Noller:2018wyv,Seraille:2024beb,Chudaykin:2024gol}.
For $p=3$ the $\gamma_{\rm B}$--$\gamma_{\rm M}$ degeneracy-direction for the Planck-alone analysis is evident, as shown at the level of the growth factor in Fig.~\ref{fig:D_contour}.
The degeneracy is broken by late-time background impact of $\gamma_{\rm M}$ with full-shape (orange) or full-shape and \textsc{DESI} (blue) in the right panel of Fig.~\ref{fig:gamma_BM_2x2}.
For $p=3/2$, the degeneracy is less visible due to the longer integrated background effect of $\gamma_{\rm M}$.
We find that adding full-shape information somewhat improves the $2\sigma$ bound on the braiding coefficient over \textsc{Planck} alone, from $\gamma_{\rm B} > -0.33$ to $\gamma_{\rm B} > -0.26$ in the $p=3/2$ case and from $\gamma_{\rm B} > -1.62$ to $\gamma_{\rm B} > -1.53$ in the $p=3$ case.

\begin{figure}[t]
    \centering
    \includegraphics[width=0.49\linewidth]{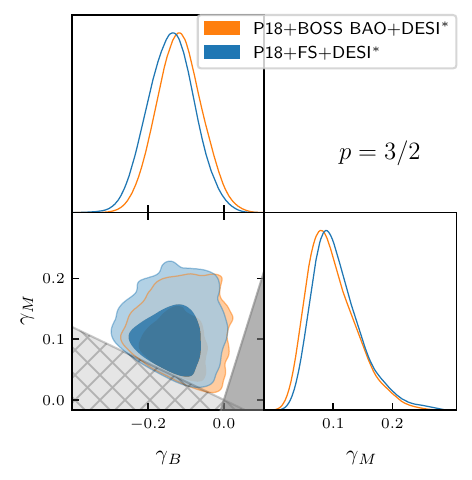}
    \includegraphics[width=0.49\linewidth]{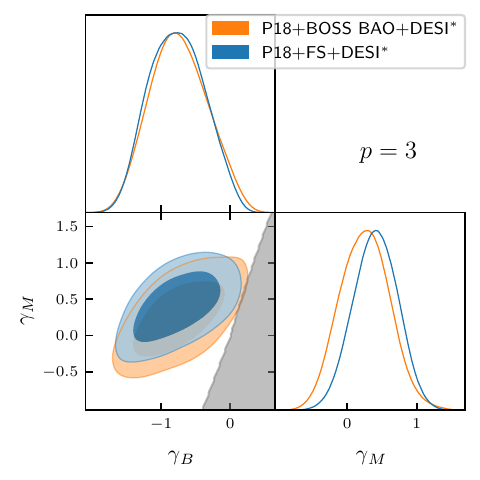}
    \caption{Same as Fig.~\ref{fig:gamma_BM_2x2}. We compare the posterior
    probabilities when exchanging \textsc{BOSS} full-shape information with \textsc{BOSS} BAO
data.}
    \label{fig:gamma_BM_2x2_bao}
\end{figure}
In Fig.~\ref{fig:gamma_BM_2x2_bao} we compare the constraining power of
using the full-shape power spectrum versus BAO measurements
(see Sec.~\ref{sec:datasets} for a description of the BAO measurements included) of BOSS
in the modified gravity model with non-zero braiding and evolving Planck mass.
We perform the comparison including \textsc{Planck} and \textsc{DESI} observations, and find
no significant improvement of using the full-shape versus BAO only within these
dataset combinations. This result emphasizes that \textsc{Planck} CMB measurements best constrain
the braiding coefficient, while the background impact dominate the constraints
on the evolution of the Planck mass.

\begin{figure}[t]
    \centering
    \includegraphics[width=0.49\linewidth]{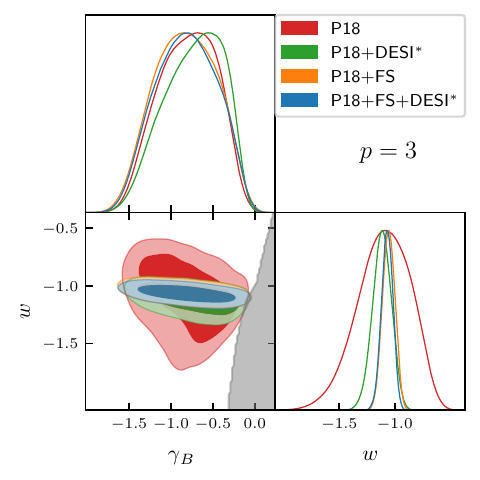}
    \includegraphics[width=0.49\linewidth]{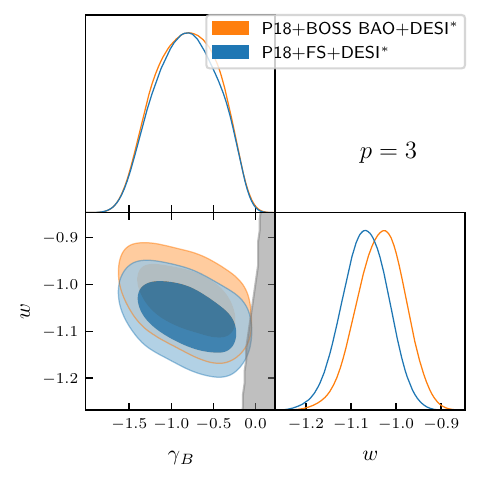}
    \caption{Posterior probabilities for Model II ($\gamma_{\rm B}$ and $\gamma_{\rm M}$
        non-zero) for $p=3$. See Sec.~\ref{sec:datasets} for definitions
        of dataset combinations. Ghost instabilities are indicated by the
        shaded area. \emph{Left:} Comparison of statistical inference with BOSS
        full-shape information and \textsc{BOSS} BAO measurements.}
    \label{fig:gammaBw0_p3_2x2}
\end{figure}

Next, we discuss the results of the statistical inference of Model II, which
considers a non-zero braiding on top of a $w$CDM background. As mentioned above, we take $p=3$ only, as
$p=3/2$ leads to tachyonic instabilities for $\gamma_{\rm B} \neq 0$. The posterior probabilities
on $\gamma_{\rm B}$ and $w$ are shown in the left panel of Fig.~\ref{fig:gammaBw0_p3_2x2}. We observe no significant
evidence for braiding effects; utilizing Planck, full-shape and \textsc{DESI} BAO data we obtain
a lower 2$\sigma$ bound $\gamma_{\rm B} > -1.46$. As for Model I, the \textsc{BOSS} full-shape power spectrum information
does not add significant additional information that can further constrain $\gamma_{\rm B}$ over \textsc{Planck} CMB data.

As is well known, \textsc{Planck} alone allows for a wide range of the dark energy equation of state $w$~\cite{Planck:2018vyg}.
The red contour in Fig.~\ref{fig:gammaBw0_p3_2x2} reproduces this conclusion, and
the full posterior in App.~\ref{app:full_posteriors} shows a strong degeneracy between $w$ and $H_0$
due to the insensitivity of CMB alone on late-time modifications. Adding late-time information from
\textsc{DESI} BAO aids in breaking the degeneracy, allowing us to measure $w = -1.12 \pm 0.17$ at 2$\sigma$.
With the full-shape
power spectrum, we can tighten the constraint on the equation of state parameter to
$w = -1.07 \pm 0.10$ at 2$\sigma$ using \textsc{Planck} + FS + DESI$^{*}$ (blue contour), showing a $40\%$ improvement. The joint posterior
probability for $\gamma_{\rm B}$ and $w$ in analyses including full-shape information (orange and blue)
displays the same degeneracy direction as shown for the growth factor in Fig.~\ref{fig:D_contour}.
This degeneracy direction is constrained by Planck's sensitivity to braiding at $\gamma_{\rm B} = -0.80^{+0.41}_{-0.32}$.

Due to the degeneracies between $H_0$ and $\gamma_{\rm M}$ and $w$ in Model I and II, respectively, the \textsc{Planck} alone analyses allows for larger values of $H_0$ compatible with the \textsc{SH0ES} measurement~\cite{Murakami:2023xuy}, see Table~\ref{tab:measurements} and full posteriors in Appendix~\ref{app:full_posteriors}.
However, adding late-time information from \textsc{BOSS} full-shape data sets breaks the degeneracy and we obtain $H_0 \simeq 69$, leaving a $\sim 3\sigma$ tension with SH0ES.%
\footnote{The combination \textsc{Planck} and \textsc{DESI}$^*$ still allows for slightly higher values of $H_0$ due to the two lowest redshift bins being removed from the \textsc{DESI} data set.}
Thus, these models fail to adequately explain the Hubble tension, consistent with the general challenges faced by late-time modifications~\cite{Schoneberg:2021qvd,Khalife:2023qbu}.
Furthermore, the models generally feature enhanced clustering at late times, thereby increasing $\sigma_8$ (see Table~\ref{tab:measurements_full}), enhancing the apparent $S_8$-tension, see e.g.~\cite{Kilo-DegreeSurvey:2023gfr}.

Lastly, we compare the full-shape constraints of Model II with an analysis using SDSS BAO measurements. Specifically, we
compare \textsc{Planck} + FS + DESI$^{*}$ to \textsc{Planck} + SDSS BAO + DESI$^{*}$. The result is shown in the right
panel of Fig.~\ref{fig:gammaBw0_p3_2x2}. As with Model I, we see similar constraining power
in both cases, with a slight shift of the posterior for $w$. Thus, we conclude that within this combination of experiments,
\textsc{Planck} is most sensitive to the braiding parameter, while the background impact of $w$ dominates
such that \textsc{BOSS} full-shape information is not significantly more sensitive to it than BAO alone.

In summary, we find in both models that \textsc{Planck} CMB measurements are mostly sensitive to
the braiding parameter $\gamma_{\rm B}$. On the other hand, the
$w$ and $\alpha_{\rm M}$ parameters are poorly constrained by CMB alone, mainly because of the degeneracy with $H_0$, see Figs.~\ref{fig:gBM_p1p5} -- \ref{fig:gBw0_p3}. LSS improves the bounds on these two parameters. Present \textsc{DESI} data do not add much information over that extracted by \textsc{BOSS} data. The comparison between FS and BAO gives similar results, meaning that the main role of  LSS is to remove the $\alpha_{\rm M}$-$H_0$ or $w$-$H_0$ degeneracies via the late-time geometrical information encoded in the BAO's.
Nonlinearities, in particular the extra dependence on $\alpha_{\rm M}$ of the non-linear effects shown in Eq.~\eqref{eq:nonlinDE} seem to play a marginal role. However, this will likely change once new FS data from \textsc{DESI} and \textsc{Euclid}, will be available.

\begin{table}[t]
    \centering
    \caption{Measured mean values and 1-$\sigma$ uncertainty on the EFTofDE parameters as well as the Hubble constant.}
    \label{tab:measurements}
    \begin{tabular}{ll|cccc}
        \toprule
                          &                             & \textbf{P18}                  & \textbf{P18 + DESI}$^{*}$     & \textbf{P18 + FS}             & \textbf{P18 + FS + DESI}$^{*}$ \\
        \midrule
        Model I           & $\gamma_{\rm B}$                  & $-0.166_{-0.083}^{+0.082}$   & $-0.146_{-0.07}^{+0.068}$    & $-0.135_{-0.065}^{+0.062}$   & $-0.140_{-0.062}^{+0.062}$  \\
        $p=3/2$           & $\gamma_{\rm M}$                  & $0.29_{-0.12}^{+0.18}$      & $0.163_{-0.097}^{+0.052}$    & $0.106_{-0.054}^{+0.029}$    & $0.106_{-0.05}^{+0.028}$    \\
                          & $H_0$                       & $73.12_{-2.2}^{+3.1}$         & $70.82_{-1.5}^{+1.0}$           & $69.77_{-0.86}^{+0.76}$       & $69.82_{-0.8}^{+0.65}$       \\
        \midrule
        Model I           & $\gamma_{\rm B}$                  & $-0.68_{-0.56}^{+0.48}$     & $-0.54_{-0.48}^{+0.44}$     & $-0.80_{-0.44}^{+0.39}$     & $-0.77_{-0.47}^{+0.37}$    \\
        $p=3$             & $\gamma_{\rm M}$                  & $0.47_{-0.71}^{+0.94}$      & $0.77_{-0.48}^{+0.55}$      & $0.31_{-0.34}^{+0.34}$      & $0.41_{-0.31}^{+0.35}$     \\
                          & $H_0$                       & $69.85_{-2.6}^{+4.4}$         & $71.16_{-1.5}^{+2.3}$         & $69.39_{-1.3}^{+1.4}$         & $69.87_{-1.1}^{+1.2}$        \\
        \midrule
        Model II          & $\gamma_{\rm B}$                  & $-0.80_{-0.32}^{+0.41}$     & $-0.73_{-0.27}^{+0.47}$     & $-0.82_{-0.40}^{+0.40}$       & $-0.82_{-0.38}^{+0.42}$    \\
        $p=3$             & $w$                         & $-1.10_{-0.22}^{+0.27}$      & $-1.116_{-0.083}^{+0.088}$    & $-1.055_{-0.057}^{+0.058}$    & $-1.069_{-0.05}^{+0.054}$    \\
                          & $H_0$                       & $71.49_{-9.7}^{+6.9}$         & $71.71_{-2.8}^{+2.4}$         & $69.7_{-1.6}^{+1.6}$          & $70.2_{-1.5}^{+1.3}$         \\
        \bottomrule
    \end{tabular}
\end{table}
The 1$\sigma$ measurements on the EFTofDE parameters are given in Table~\ref{tab:measurements}, while the full list of all cosmological parameters is given in Table~\ref{tab:measurements_full}.

\section{Conclusion}
\label{sec:Conclusion}
We have performed a combined analysis of CMB and LSS data, including the recent BAO measurement from the \textsc{DESI} collaboration, constraining modified gravity and dark energy properties using the EFTofDE framework. The main new features of our pipeline are a consistent treatment of the linear background equations, as discussed in Sec.~\ref{sec:BK}, and a consistent treatment of dark matter nonlinearities, see Sec.~\ref{sec:DM}.

We have shown that the combination with LSS data can improve the constraints on the EFTofDE parameter space compared to CMB alone analyses.
This is true for the  modified gravity parameter $\gamma_{\rm M}$, which quantifies the running of the Planck mass, and the equation of state of dark energy $w$. The main information provided by current LSS observations is contained in BAO's, which allow to break the geometrical degeneracy between the Hubble rate and parameters that affect the background evolution, either the dark energy equation of state or the running of the Planck mass.
One goal of this work was to assess at which level the remaining information encoded in the power spectrum shape is accessible, in particular from nonlinear scales.
For the \textsc{BOSS} data and covariances we do not detect a significant improvement in the constraints on the EFTofDE parameters.
This becomes clear when looking to the constraints on the braiding parameter $\gamma_{\rm B}$, which affects the perturbations without altering the background evolution: this parameter is mostly constrained by CMB.
Adding LSS information, either BAO alone or full-shape power spectrum, has a negligible impact.
However, future full shape data from \textsc{Euclid} and \textsc{DESI} will likely change the situation, given their large observed volumes and number densities, thus requiring consistent nonlinear pipelines for their analyses based on the EFTofLSS, as the one developed in this work. Moreover, the wide range of redshifts that will be observed will definitely improve our understanding of background evolution, shedding light on dark energy dynamics. 

Including higher-order correlation functions such as the bispectrum is not expected to significantly ameliorate the current situation, given its low signal-to-noise ratio. It has been shown that the main contribution from the bispectrum is an improved estimate for the non-linear bias parameters and a $\sim 5-15\%$ improvement on cosmological parameters assuming $\Lambda$CDM~\cite{Ivanov:2021kcd, DAmico:2022osl, DAmico:2022ukl}. Even for beyond-$\Lambda$CDM models, such as including new dark matter species or non-standard dark energy dynamics, adding the LSS bispectrum does not significantly improve the constraints~\cite{Tsedrik:2022cri, Rogers:2023ezo}.

Moreover, in the EFTofDE, parameters appear at the level of the action, that is, of the equations of motion. In order to solve the equations of motion and derive the observables, a time-dependence of these functions needs to be assumed. In this paper we tested two different power laws, see Eq.~\eqref{Gammapar}. The corresponding constraints are shown for instance in Fig.~\ref{fig:gamma_BM_2x2}, which shows that these two choices correspond to two very different models. Therefore, the question arises of how to treat the model dependence hidden in the time-dependence. A possible, agnostic, alternative would be to parameterize directly the observables, in place of the equations of motion, exploiting symmetries, as in the bootstrap approach of Refs.~\cite{DAmico:2021rdb, Marinucci:2024add}. This would require introducing new coefficients parameterizing the ISW and lensing effects and a set of new parameters for any redshift bin of LSS data.
The design of strategies to optimize the trade-off between model-independence and parameter proliferation is therefore a crucial issue in the exploration of physics beyond $\Lambda$CDM.
Most of the information loss in LSS arises due to the ignorance about the non-linear mapping from dark matter to galaxies parameterized by the galaxy biases.
One possible way to tackle this is putting physically motivated priors on them, using e.g.\ analytical recipes~\cite{Kaiser:1984sw,Bardeen:1985tr, Desjacques:2010gz,Musso:2012qk, Desjacques:2016bnm,Marinucci:2019wdb, Marinucci:2020weg}, or numerical simulations~\cite{Ivanov:2024hgq}.

We leave all these possible extensions of our work to future studies.

\vspace{0.5cm}


\noindent {\bf Acknowledgments:} We are grateful to Emilio Bellini, Guido D'Amico, Lorenzo Piga and Anton Chudaykin for useful discussions related to this work.
MM and MP acknowledge support by the MIUR Progetti di Ricerca di Rilevante Interesse Nazionale (PRIN) Bando 2022 - grant 20228RMX4A, funded by the European Union - Next generation EU, Mission 4, Component 1, CUP C53D23000940006.
PT and FV acknowledge  support by the ANR Project COLSS (ANR-21-CE31-0029).
This work was granted access to the CCRT High-Performance Computing (HPC) facility under the Grant CCRT2024-taulepet awarded by the Fundamental Research Division (DRF) of CEA.
The package GetDist~\cite{Lewis:2019xzd} was used to create the plots and compute the summary statistics.

\newpage
\appendix

\section{Tensor speed excess}
\label{app:alphaT}

\begin{figure}
    \centering
    \includegraphics[width=0.48\linewidth]{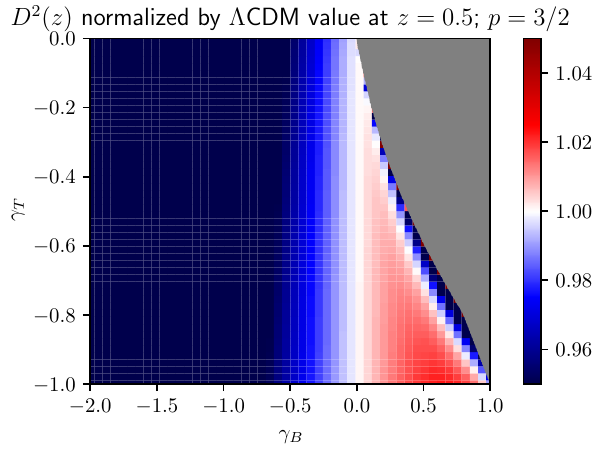}
    \includegraphics[width=0.48\linewidth]{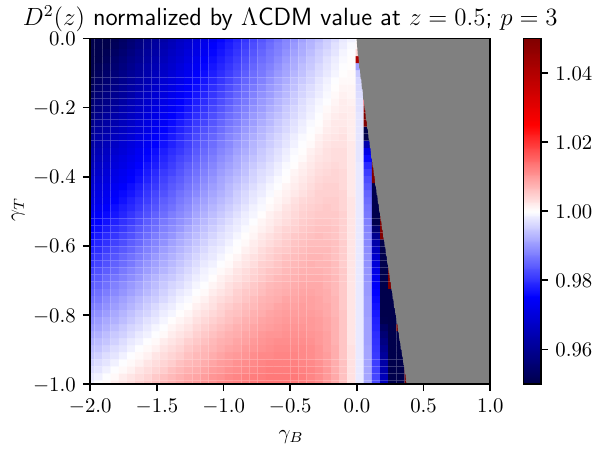}
    \caption{Growth factor $D^2$ at $z=0.5$ normalized to the $\Lambda$CDM value in models with non-zero braiding and tensor speed excess ($\Lambda$CDM background).
    The gray shaded regions indicate ghost instabilities.
    \emph{Left}: $p=3/2$ and \emph{right}: $p=3$.
    }
    \label{fig:gammaBT_contour}
\end{figure}
In this appendix we consider the impact of tensor speed excess $\alpha_{\rm T}$ on LSS.
We set $\alpha_{\rm M} = 0$ and $w=-1$ for simplicity and choose $\alpha_{\rm T} = \gamma_{\rm T} (a/a_0)^p$. Then the function $\nu$ (see Eq.~\eqref{nu}) is given by
\be
 \nu = - (1+ \alpha_{\rm B})\left[\alpha_{\rm B} + \alpha_{\rm T}(1+\alpha_{\rm B})  \right]  + \left(  \frac32 \Omega_{\rm m} -p  \right) \alpha_{\rm B}  \,,
\label{eq:alphaT_nu}
\ee
and moreover we obtain
\be
 \mu_{\Phi} = 1 + \alpha_{\rm T} + \frac{\left[\alpha_{\rm T} + \alpha_{\rm B}(1 + \alpha_{\rm T})\right]^2}{\nu} \,.
\label{eq:alphaT_mu}
\ee
When $p=3/2$, the last parenthesis of Eq.~\eqref{eq:alphaT_nu} is close to zero until very late times, resulting in a cancellation of $\alpha_{\rm T}$ at leading order in $\mu_{\Phi}$.
Thus, at leading order the Poisson equation is unaltered by the tensor speed excess, as can be seen in the left panel of Fig.~\ref{fig:gammaBT_contour}, which displays the growth factor $D^2(z)$ at $z=0.5$ in this model.
For $p=3$, we can instead identify a degeneracy direction in which $\mu_{\Phi} = 1$:
\begin{equation}
    \alpha_{\rm T} = - \frac{2 \alpha_{\rm B}}{-4 + 2 \alpha_{\rm B} + 3 \Omega_{\rm m}}\,.
    \label{eq:alpha_T_degeneracy}
\end{equation}
This leads to a region of parameter space which is unaffected by the modified gravity parameters, as exemplified by the right panel of Fig.~\ref{fig:gammaBT_contour}.
(Note that due to the growth factor being an integrated quantity of $\mu_{\Phi}$ and the $\alpha$s and $\Omega_{\rm m}$ having different time-dependencies, the degeneracy in $D$ does not simply have the functional form of Eq.~\eqref{eq:alpha_T_degeneracy}.)

In conclusion, for the parameterizations considered here, the tensor speed excess impact on the Poisson equation is too small for LSS data to be sensitive to it.

\section{Full parameter constraints}
\label{app:full_posteriors}

\begin{figure}[p]
    \includegraphics[width=1\linewidth]{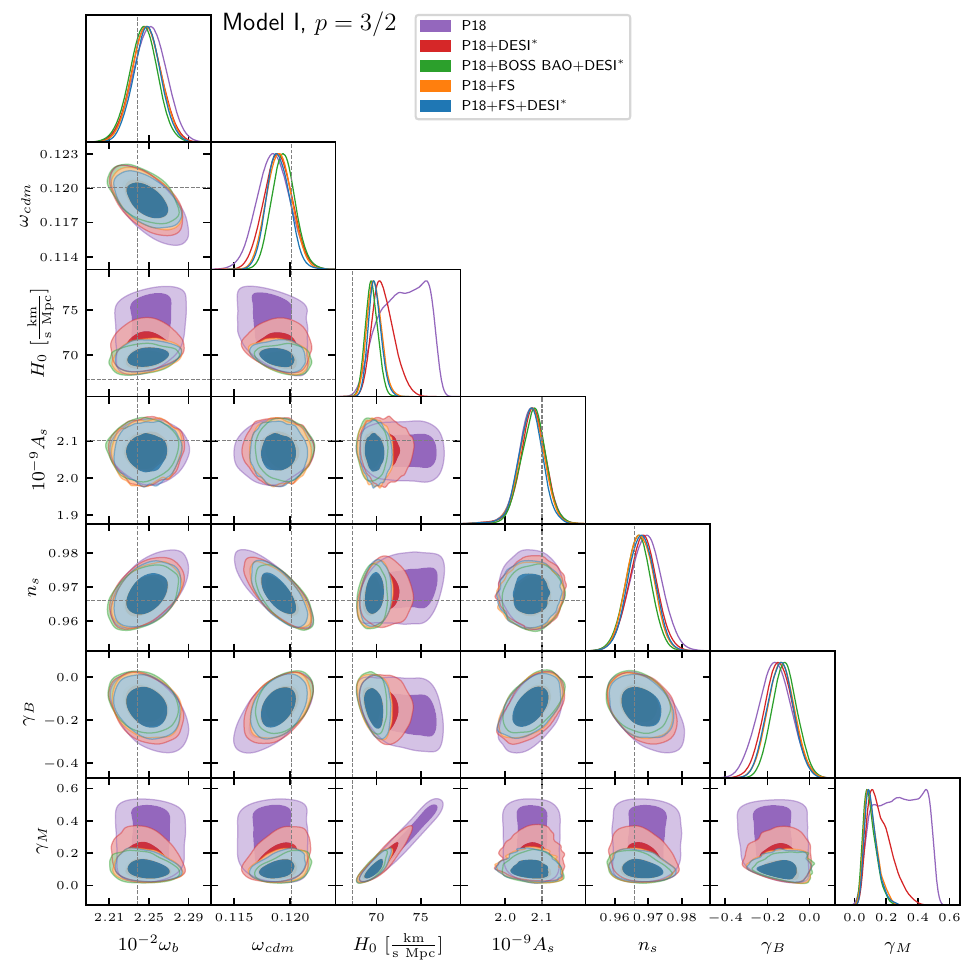}
    \caption{Full posterior probability triangle plot for Model I with $p=3/2$. The dashed lines indicate the best-fit values in $\Lambda$CDM for the P18 data set~\cite{Planck:2018vyg}.}
    \label{fig:gBM_p1p5}
\end{figure}

\begin{figure}[p]
    \includegraphics[width=1\linewidth]{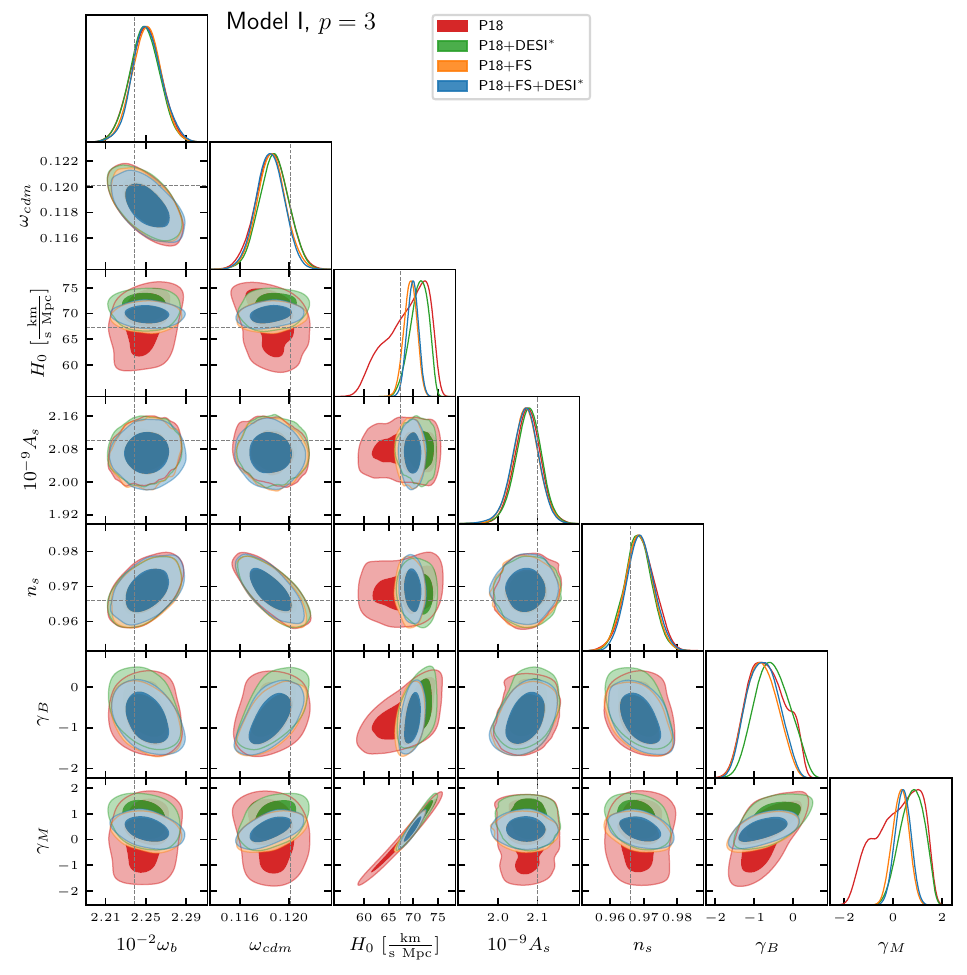}
    \caption{Full posterior probability triangle plot for Model I with $p=3$. The dashed lines indicate the best-fit values in $\Lambda$CDM for the P18 data set~\cite{Planck:2018vyg}.}
    \label{fig:gBM_p3}
\end{figure}

\begin{figure}[p]
    \includegraphics[width=1\linewidth]{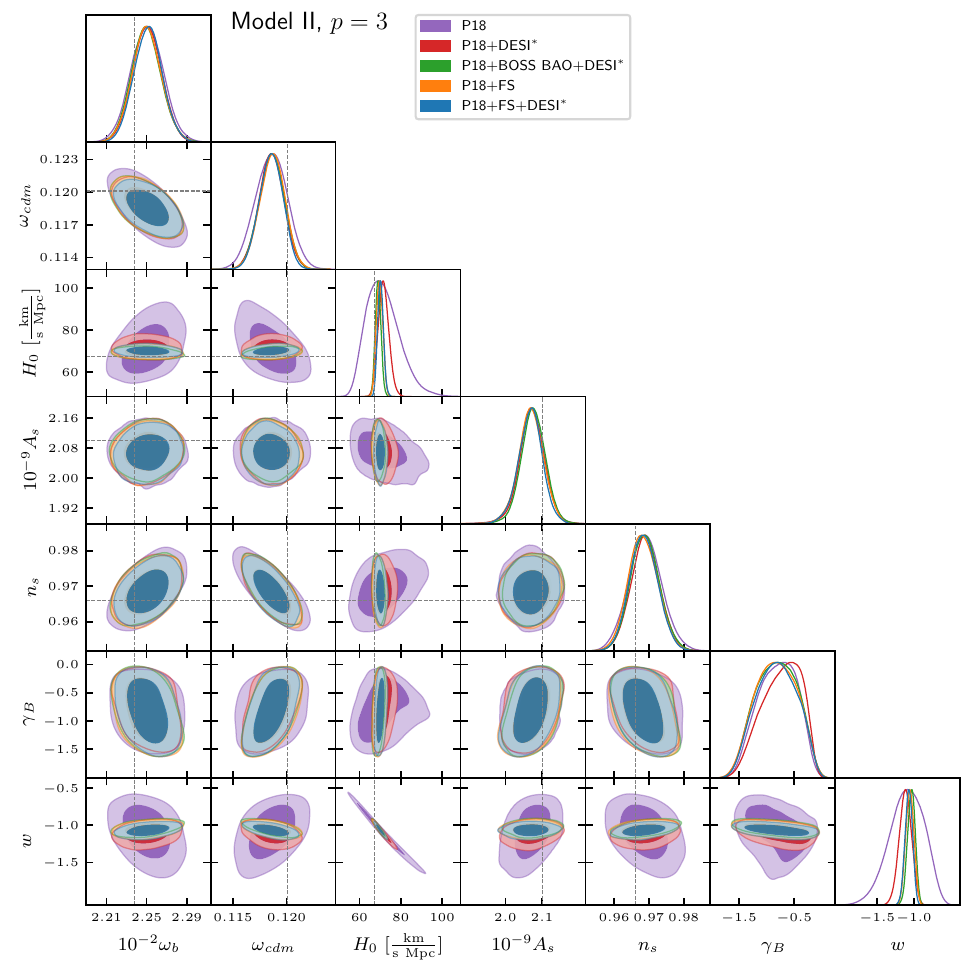}
    \caption{Full posterior probability triangle plot for Model II with $p=3$. The dashed lines indicate the best-fit values in $\Lambda$CDM for the P18 data set~\cite{Planck:2018vyg}.}
    \label{fig:gBw0_p3}
\end{figure}

We display triangle plots of the standard cosmological parameters and the EFTofDE parameters in Fig.~\ref{fig:gBM_p1p5} (Model I with $p=3/2$), Fig.~\ref{fig:gBM_p3} (Model I with $p=3$) and Fig.~\ref{fig:gBw0_p3} (Model II with $p=3$). The measured mean and 1$\sigma$ on the parameters are reported in Table~\ref{tab:measurements_full}.
\begin{table}[p]
    \centering
    \caption{The inferred mean and 1$\sigma$ uncertainty on all cosmological parameters. We quote also the maximum likelihood.}
    \label{tab:measurements_full}
    \begin{tabular}{ll|cccc}
        \toprule
                          &                             & \textbf{P18}                  & \textbf{P18 + DESI}$^{*}$     & \textbf{P18 + FS}             & \textbf{P18 + FS + DESI}$^{*}$ \\
        \midrule
        Model I           & $100\omega_b$               &  $2.251_{-0.017}^{+0.018}$    & $2.247_{-0.016}^{+0.015}$     & $2.248_{-0.015}^{+0.014}$     & $2.248_{-0.013}^{+0.014}$    \\
        $p=3/2$           & $\omega_{cdm}$              &  $0.1185_{-0.0016}^{+0.0016}$ & $0.1189_{-0.0013}^{+0.0013}$  & $0.119_{-0.0012}^{+0.0011}$   & $0.1189_{-0.0011}^{+0.001}$  \\
                          & $H_0~[\frac{\mathrm{km}}{\mathrm{s}\,\mathrm{Mpc}}]$        & $73.12_{-2.2}^{+3.1}$         & $70.82_{-1.5}^{+1}$           & $69.77_{-0.86}^{+0.76}$       & $69.82_{-0.8}^{+0.65}$ \\
                          & $10^{+9}A_s$                &  $2.072_{-0.038}^{+0.039}$    & $2.075_{-0.011}^{+0.13}$      & $2.071_{-0.062}^{+0.049}$     & $2.068_{-0.08}^{+0.069}$     \\
                          & $\sigma_8$                  &  $0.891_{-0.033}^{+0.042}$    & $0.8616_{-0.048}^{+0.024}$    & $0.8462_{-0.023}^{+0.021}$    & $0.8459_{-0.033}^{+0.015}$   \\
                          & $n_s$                       & $0.9693_{-0.0051}^{+0.005}$   & $0.9684_{-0.0042}^{+0.0044}$  & $0.9676_{-0.0042}^{+0.0042}$  & $0.968_{-0.0042}^{+0.004}$   \\
                          & $\tau_{\mathrm{reio}}$      & $0.0498_{-0.0086}^{+0.0093}$  & $0.04983_{-0.0077}^{+0.0089}$ & $0.04875_{-0.0078}^{+0.0084}$ & $0.04827_{-0.0073}^{+0.0085}$\\
                          & $\gamma_{\rm B}$            & $-0.1655_{-0.083}^{+0.082}$   & $-0.1464_{-0.07}^{+0.068}$    & $-0.1351_{-0.065}^{+0.062}$   & $-0.1399_{-0.062}^{+0.062}$  \\
                          & $\gamma_{\rm M}$            & $0.2865_{-0.12}^{+0.18}$      & $0.1629_{-0.097}^{+0.052}$    & $0.1061_{-0.054}^{+0.029}$    & $0.1061_{-0.05}^{+0.028}$    \\
                          & $-2\ln{\cal L}_\mathrm{min}$& $2777.2$                      & $2783.4$                      & $2921.9$                      & $2928.9$                     \\
        \midrule
        Model I           & $100\omega_b$               & $2.249_{-0.02}^{+0.017}$      & $2.249_{-0.015}^{+0.015}$     & $2.251_{-0.014}^{+0.015}$     & $2.251_{-0.016}^{+0.014}$    \\
        $p=3$             & $\omega_{cdm}$              & $0.1187_{-0.0016}^{+0.0016}$  & $0.1188_{-0.0012}^{+0.0013}$  & $0.1186_{-0.0012}^{+0.0012}$  & $0.1185_{-0.0012}^{+0.0011}$ \\
                          & $H_0~[\frac{\mathrm{km}}{\mathrm{s}\,\mathrm{Mpc}}]$        & $69.85_{-2.6}^{+4.4}$         & $71.16_{-1.5}^{+2.3}$         & $69.39_{-1.3}^{+1.4}$         & $69.87_{-1.1}^{+1.2}$ \\
                          & $10^{+9}A_s$                & $2.075_{-0.039}^{+0.14}$      & $2.077_{-0.075}^{+0.039}$     & $2.07_{-0.076}^{+0.039}$      & $2.069_{-0.033}^{+0.034}$    \\
                          & $\sigma_8$                  & $0.8447_{-0.091}^{+0.064}$    & $0.8609_{-0.041}^{+0.04}$     & $0.8332_{-0.036}^{+0.022}$    & $0.8382_{-0.021}^{+0.019}$   \\
                          & $n_s$                       & $0.9684_{-0.0052}^{+0.0053}$  & $0.9683_{-0.004}^{+0.0042}$   & $0.9685_{-0.0041}^{+0.0043}$  & $0.969_{-0.0044}^{+0.0039}$  \\
                          & $\tau_{\mathrm{reio}}$      & $0.05016_{-0.0089}^{+0.0094}$ & $0.05062_{-0.0073}^{+0.0083}$ & $0.04934_{-0.0075}^{+0.0081}$ & $0.04913_{-0.0074}^{+0.0084}$\\
                          & $\gamma_{\rm B}$            & $-0.6751_{-0.56}^{+0.48}$     & $-0.5371_{-0.48}^{+0.44}$     & $-0.8006_{-0.44}^{+0.39}$     & $-0.7725_{-0.47}^{+0.37}$    \\
                          & $\gamma_{\rm M}$            & $0.4727_{-0.71}^{+0.94}$      & $0.7675_{-0.48}^{+0.55}$      & $0.3096_{-0.34}^{+0.34}$      & $0.4096_{-0.31}^{+0.35}$     \\
                          & $-2\ln{\cal L}_\mathrm{min}$& $2776.6$                      & $2781.8$                      & $2921.7$                      & $2927.8$ \\
        \midrule
        Model II          & $100\omega_b$               & $2.251_{-0.017}^{+0.015}$     & $2.251_{-0.015}^{+0.014}$     & $2.25_{-0.014}^{+0.015}$      & $2.252_{-0.015}^{+0.014}$    \\
        $p=3$             & $\omega_{cdm}$              & $0.1186_{-0.0015}^{+0.0016}$  & $0.1186_{-0.0011}^{+0.0012}$  & $0.1187_{-0.0011}^{+0.0012}$  & $0.1186_{-0.0011}^{+0.0012}$ \\
                          & $H_0~[\frac{\mathrm{km}}{\mathrm{s}\,\mathrm{Mpc}}]$        & $71.49_{-9.7}^{+6.9}$         & $71.71_{-2.8}^{+2.4}$         & $69.7_{-1.6}^{+1.6}$          & $70.2_{-1.5}^{+1.3}$  \\
                          & $10^{+9}A_s$                & $2.071_{-0.078}^{+0.041}$     & $2.072_{-0.033}^{+0.035}$     & $2.069_{-0.033}^{+0.035}$     & $2.067_{-0.033}^{+0.033}$    \\
                          & $\sigma_8$                  & $0.8437_{-0.08}^{+0.065}$     & $0.8482_{-0.026}^{+0.023}$    & $0.8308_{-0.017}^{+0.016}$    & $0.8341_{-0.015}^{+0.015}$  \\
                          & $n_s$                       & $0.9689_{-0.0048}^{+0.0048}$  & $0.969_{-0.0041}^{+0.0042}$   & $0.9683_{-0.0043}^{+0.0042}$  & $0.9685_{-0.0042}^{+0.0039}$ \\
                          & $\tau_{\mathrm{reio}}$      & $0.04946_{-0.0076}^{+0.0086}$ & $0.04986_{-0.0079}^{+0.0084}$ & $0.04901_{-0.0076}^{+0.0083}$ & $0.04886_{-0.0073}^{+0.0081}$\\
                          & $\gamma_{\rm B}$            & $-0.7992_{-0.32}^{+0.41}$     & $-0.7273_{-0.27}^{+0.47}$     & $-0.8228_{-0.4}^{+0.4}$       & $-0.8176_{-0.38}^{+0.42}$    \\
                          & $w$                         & $-1.098_{-0.22}^{+0.27}$      & $-1.116_{-0.083}^{+0.088}$    & $-1.055_{-0.057}^{+0.058}$    & $-1.069_{-0.05}^{+0.054}$    \\
                          & $-2\ln{\cal L}_\mathrm{min}$& $2775.7$                      & $2780.6$                      & $2921.2$                      & $2927.4$ \\
        \bottomrule
    \end{tabular}
\end{table}
\newpage

\providecommand{\href}[2]{#2}\begingroup\raggedright\endgroup

\end{document}